\documentclass[runningheads,a4paper]{aa}    
\usepackage{graphicx,url,twoopt,natbib}
\usepackage{multirow}
\usepackage[varg]{txfonts}           
\usepackage{hyperref}                
\usepackage{xcolor}

\hypersetup{
  colorlinks=true,   
  urlcolor=blue,     
  linkcolor=blue     
}

\bibpunct{(}{)}{;}{a}{}{,}    
\graphicspath{ {figures/} }

\makeatletter
\newcommand{\bibnote}[2]{\global\@namedef{#1note}{#2}}
\newcommand{\biblink}[2]{\global\@namedef{#1link}{#2}}
\newcommand{\Alfven}[1]{Alfv\'en}

\newcommand{\fancyE}{\mathcal{E}}

\makeatother

\makeatletter
  \protected\def\stonyslink{%
     \def\hyper@linkstart##1##2{}\let\hyper@linkend\@empty}
  \newcommandtwoopt{\citeads}[3][][]{%
   \href{http://ui.adsabs.harvard.edu/abs/#3/abstract}%
        {\stonyslink \citealp[#1][#2]{#3}}
   \biblink{#3}{\href{http://ui.adsabs.harvard.edu/abs/#3/abstract}{ADS}}}
 \newcommandtwoopt{\citepads}[3][][]{%
   \href{http://ui.adsabs.harvard.edu/abs/#3/abstract}%
        {\stonyslink \citep[#1][#2]{#3}}
   \biblink{#3}{\href{http://ui.adsabs.harvard.edu/abs/#3/abstract}{ADS}}}
 \newcommandtwoopt{\citetads}[3][][]{%
   \href{http://ui.adsabs.harvard.edu/abs/#3/abstract}%
        {\stonyslink \citet[#1][#2]{#3}}
  \biblink{#3}{\href{http://ui.adsabs.harvard.edu/abs/#3/abstract}{ADS}}}
 \newcommandtwoopt{\citeyearads}[3][][]{%
   \href{http://ui.adsabs.harvard.edu/abs/#3/abstract}%
        {\stonyslink \citeyear[#1][#2]{#3}}
   \biblink{#3}{\href{http://ui.adsabs.harvard.edu/abs/#3/abstract}{ADS}}}
\makeatother

\begin{document}  


   \title{Modelling the interaction of Alfv\'enic fluctuations with coronal mass ejections in the low solar corona}
   \titlerunning{The interaction of Alfv\'enic fluctuations with CMEs}

 \author{Chaitanya Prasad Sishtla\inst{1} \and
          Jens Pomoell\inst{1} \and
          Rami Vainio\inst{2} \and
          Emilia Kilpua\inst{1} \and
          Simon Good\inst{1}
          }

  \institute{Department of Physics, University of Helsinki,
              Helsinki, Finland\\
              \email{chaitanya.sishtla@helsinki.fi}
              \and
              Department of Physics \& Astronomy, University of Turku,
              Turku, Finland
 }

   \date{}

 
  \abstract
   {Alfv\'enic fluctuations of various scales are ubiquitous in the corona;  their non-linear interactions and eventual turbulent cascade result in an important heating mechanism that accelerates the solar wind. These fluctuations may be processed by large-scale, transient, and coherent heliospheric structures such as coronal mass ejections (CMEs).  
   In this study we investigate the interactions between Alfv\'enic solar wind fluctuations and CMEs using magnetohydrodynamic (MHD) simulations.}
   {We study the transmission of upstream solar wind fluctuations into the CME leading to the formation of CME sheath fluctuations. Additionally, we investigate the influence of the fluctuation frequencies on the extent of the CME sheath.}
   {We used an ideal  MHD model with an adiabatic equation of state. An Alfv\'en pump wave is injected into the quiet solar wind by perturbing the transverse magnetic field and velocity components, and a CME is injected by inserting a flux-rope modelled as a magnetic island into the quasi-steady solar wind.}
   {The upstream Alfv\'en waves experience a decrease in wavelength and change in the wave vector direction due to the non-radial topology of the CME shock front. The CME sheath inhibits the transmission of long-wavelength fluctuations due to the presence of non-radial flows in this region. The frequency of the solar wind fluctuations also affects the steepening of MHD fast waves causing the CME shock propagation speed to vary with the solar wind fluctuation frequencies.}
   {}
   

   \maketitle
%
\section{Introduction}     \label{sec:introduction}
The turbulent fluctuations in velocity, magnetic field, electric field, and density are ubiquitous in the solar wind and corona~\citep{coleman1968turbulence, belcher1971large, bale2005measurement}. The convective motions of the dense photospheric plasma, which contains the solar magnetic field, are considered to be the primary source of energy for these fluctuations~\citep{cranmer2005generation, kato2016chromospheric}. These fluctuations have been observed both in situ~\citep{belcher1971large, dAmicis2015} and remotely~\citep{tomczyk2007alfven}. In the solar wind, the power contained in Alfv\'enically polarised fluctuations dominates over the power in compressive fluctuations~\citep{tu1995mhd, chen2016recent}. Additionally, the solar wind exhibits broad-band Alfv\'enic fluctuations, which can then non-linearly interact to initiate an energy cascade leading to dissipation via heating on smaller spatial scales. In this view of a turbulence cascade, the inertial range is the spatial scale of the fluctuations exhibiting a power-law behaviour between the energy injection and dissipation scales. This inertial-range turbulence is often studied within the framework of reduced magnetohydrodynamics (RMHD), in which Alfv\'en waves are the linear wave modes~\citep{zank1992equations, schekochihin2009astrophysical, perez2013direct}. Previous studies~\citep{matthaeus1984particle, gershman2019alfvenic, gonzalez2021proton} have also discussed the role of Alfv\'en wave propagation and reflection-driven Alfv\'enic turbulence for particle acceleration in planetary radiation belts, in MHD reconnection sites, and at interplanetary discontinuities.

In this study we investigate the interaction of Alfv\'enic perturbations with a coronal mass ejection (CME) in the low corona. A CME is a transient plasma and magnetic field eruption from the solar corona; it exhibits complex magnetic substructures. They are one of the primary drivers of geomagnetic activity near Earth~\citep{kilpua2013magnetic, kilpua2015unraveling, kalliokoski2020outer, kalliokoski2022outer}. In coronagraph images CMEs often exhibit a three-part structure with a bright loop of compressed coronal plasma enclosing a dark low-density cavity, corresponding to a flux rope (FR), which contains a high-density core~\citep{gibson2000three, kilpua2017coronal}. A spacecraft encountering a CME typically observes a shock, followed by a turbulent sheath and the ejecta. Only part of the ejecta at 1 AU shows clear FR signatures due to interaction and evolution. The internal structure of the CME is of significant interest as FR can cause strong and sustained southward magnetic fields influencing the Earth~\citep{kilpua2017geoeffective}. In addition, the turbulent and compressed CME sheath is highly geoeffective~\citep{kilpua2017coronal, kilpua2019solar}. Sheaths are known to exhibit an extensive range of inertial and kinetic range spectral indices~\citep{kilpua2020magnetic, kilpua2021statistical}, embed multi-scale structures~\citep{ruohotie2022fluxrope}, and contribute to the acceleration of solar energetic particles~\citep{kilpuaSEP}. 
The fluctuations in the CME sheath have been seen to exhibit turbulence characteristics often observed in the slow solar wind (e.g. higher compressibility), and yet they are still dominated by non-compressible Alfv\'enic fluctuations~\citep{moissard2019study}. Additionally, compared to the predominantly anti-sunward fluctuations in the solar wind preceding CMEs near 1 AU, sheaths are found to exhibit a more balanced distribution of sunward and anti-sunward fluctuations~\citep{good2020cross, good2022cross, soljento2023imbalanced}.
There is currently a wide range of models~\citep{gibson1998time, isavnin2016fried, verbeke2019spheromak, asvestari2021modelling, asvestari2022spheromak} to investigate the evolution of the global FR structure of CMEs from Sun to Earth and their interactions with the ambient solar wind. However, we need the understanding and capabilities to model the smaller-scale features of CMEs and their sheath regions. One important aspect is the transmission of Alfv\'enic fluctuations from the surrounding ambient solar wind into the CME and the role it plays in forming the sheath. 

In this study, by using numerical simulations, we aim to enhance our understanding of the formation of sheath structures by demonstrating the effect of Alfv\'enic solar wind fluctuations on the large-scale structures of the CME and to analyse the transmission of these fluctuations into the sheath region. We find the CME shock speed influenced by the frequency of solar wind fluctuations, with the CME sheath exhibiting non-radial flows, along with both sunward and anti-sunward Alfv\'enic fluctuations. These results are obtained by performing 2.5D MHD simulations of the solar corona assuming a radial solar magnetic field, with the FR modelled using the Grad-Shafranov equation.

In Section~\ref{sec:methodology} we introduce the MHD equations and associated boundary conditions, the mechanism for Alfv\'en wave injection, and the CME model used for the simulations. The influence of solar wind fluctuations on the CME and their transmission to the sheath is discussed in Section~\ref{sec:results}. Section~\ref{sec:cme-sheath} presents a statistical comparison of the shock location and sheath extent for varying solar wind and CME parameters, including a case with no solar wind fluctuations. The conclusions are summarised in Section~\ref{sec:conclusion}.

\section{Methodology}     \label{sec:methodology}
To perform our study we developed a 2.5D magnetohydrodynamic (MHD) simulation from the low corona at $1.03$ solar radii ($R_\odot$) to $30~R_\odot$. The simulation domain is 2D in space;    the velocity and electromagnetic field vectors have three components. The solar wind is modelled assuming a global radial unipolar (outward) solar magnetic field, which can be considered realistic for a limited region of the Sun, such as a coronal hole. The MHD equations and the relevant physical processes of gravity and ad hoc coronal heating are described by the following equations:

\begin{align}
    \frac{\partial \rho}{\partial t} + \nabla \cdot (\rho \mathbf{v}) = 0,
\label{eq:mass-cont}
\end{align}
\begin{align}
    \frac{\partial (\rho\mathbf{v})}{\partial t} + \nabla \cdot [\rho\mathbf{v}\mathbf{v} + (P+\frac{B^2}{2\mu_0})\mathbf{I} - \frac{\mathbf{B}\mathbf{B}}{\mu_0}] = -\frac{GM_\odot\rho}{r^2}\hat{\mathbf{r}},
\label{eq:mom-cont}
\end{align}
\begin{align}
    \frac{\partial \fancyE}{\partial t} + \nabla\cdot[(\fancyE + P - \frac{B^2}{2\mu_0})\mathbf{v} + \frac{1}{\mu_0}\mathbf{B}\times(\mathbf{v}\times\mathbf{B})] = -\frac{GM_\odot\rho v_r}{r^2} + S,
\label{eq:energy-cont}
\end{align}
\begin{align}
    \nabla \cdot \mathbf{B} = 0,
\label{eq:gauss-law}
\end{align}
\begin{align}
    \frac{\partial \mathbf{B}}{\partial t} - \nabla \times (\mathbf{v}\times\mathbf{B}) = 0,
\label{eq:induction}
\end{align}
where
\begin{align}
    \fancyE = \frac{1}{2}\rho v^2 + \frac{P}{\gamma-1} + \frac{B^2}{2\mu_0},
\label{eq:def:energy}
\end{align}
\begin{align}
    S = S_0 \mathrm{exp}\left(-\frac{r}{L}\right).
\label{eq:def:exp-heating}
\end{align}
Here the quantities $\rho$, $\mathbf{v}$, $\mathbf{B}$, $\fancyE$, and $P$ correspond to the mass density, bulk plasma velocity, magnetic field, total energy density, and thermal pressure. Equations~\ref{eq:mass-cont}-\ref{eq:induction} correspond to the mass continuity, momentum, energy continuity, and induction equations, respectively, and are subsequently   referred to as such. The solar wind plasma evolves by solving these MHD equations for an adiabatic polytropic index of $\gamma = 5/3$. Thus, to obtain a steady-state solar wind that approximates a Parker-like outflow, we incorporate an additional energy source term in Equation~\ref{eq:def:exp-heating}~\citep{pomoell2015modelling, mikic2018predicting, sishtla2022flux} with $S_0 = 0.5\times 10^{-6}$~\textrm{Wm}$^{-3}$ and $L = 0.4~R_\odot$.

The numerical method used in this work to solve the MHD equations was employed in previous studies of the solar corona~\citep{pomoell2012influence}. The method utilises a strong stability preserving (SSP) Runge-Kutta method to advance the semi-discretised equations in time, and employs the Harten–Lax–van Leer (HLL) approximate Riemann solver supplied by piece-wise linear slope-limited interface states. The equations are solved in spherical coordinates and the magnetic field is ensured to be divergence free to the floating point accuracy by utilising the constrained transport method~\citep{kissmann2012semidiscrete}. 

The MHD equations were integrated forward in time for a 2D meridional plane with a radial extent of $r = r_0 = 1.03~R_\odot$ to $r = 30~R_\odot$, and an co-latitudinal extent of $\theta = 10^\circ$ to $\theta = 170^\circ$. The domain is therefore symmetric in the out-of-plane longitudinal $\phi$ direction. The solar magnetic field was initialised to be radially outward with an associated vector potential $\mathbf{A} = -B_0 r_0 (r_0/r)\cot{\theta}~\hat{\mathbf{\phi}}$, where $B_0 = 5$~\textrm{G}, and the magnetic field in the simulation was then specified using $\mathbf{B = \nabla\times A}$. The simulation grid was defined by $500$ cells logarithmically spaced in the radial direction, and $128$ equidistant cells in the latitudinal direction. Appendix~\ref{appendix:shock-compression-aw} validates this choice of the radial grid resolution by verifying the results presented in the following sections for a significantly higher resolution. 

At the inner radial boundary, representing the coronal base,  we specified a constant mass density and temperature along the boundary with $\rho_0 = 8.5\times 10^{-13}~\mathrm{kg}$ and $T_0 = 1.2\times 10^6~\mathrm{K}$. At the outer radial and the latitudinal boundaries, we linearly extrapolated all the  dynamical quantities to enforce an outflow boundary condition.

\subsection{Introducing Alfv\'enic perturbations} \label{subsec:methods/AW}
\begin{figure}[ht]
\centering
\includegraphics[width=0.5\textwidth]{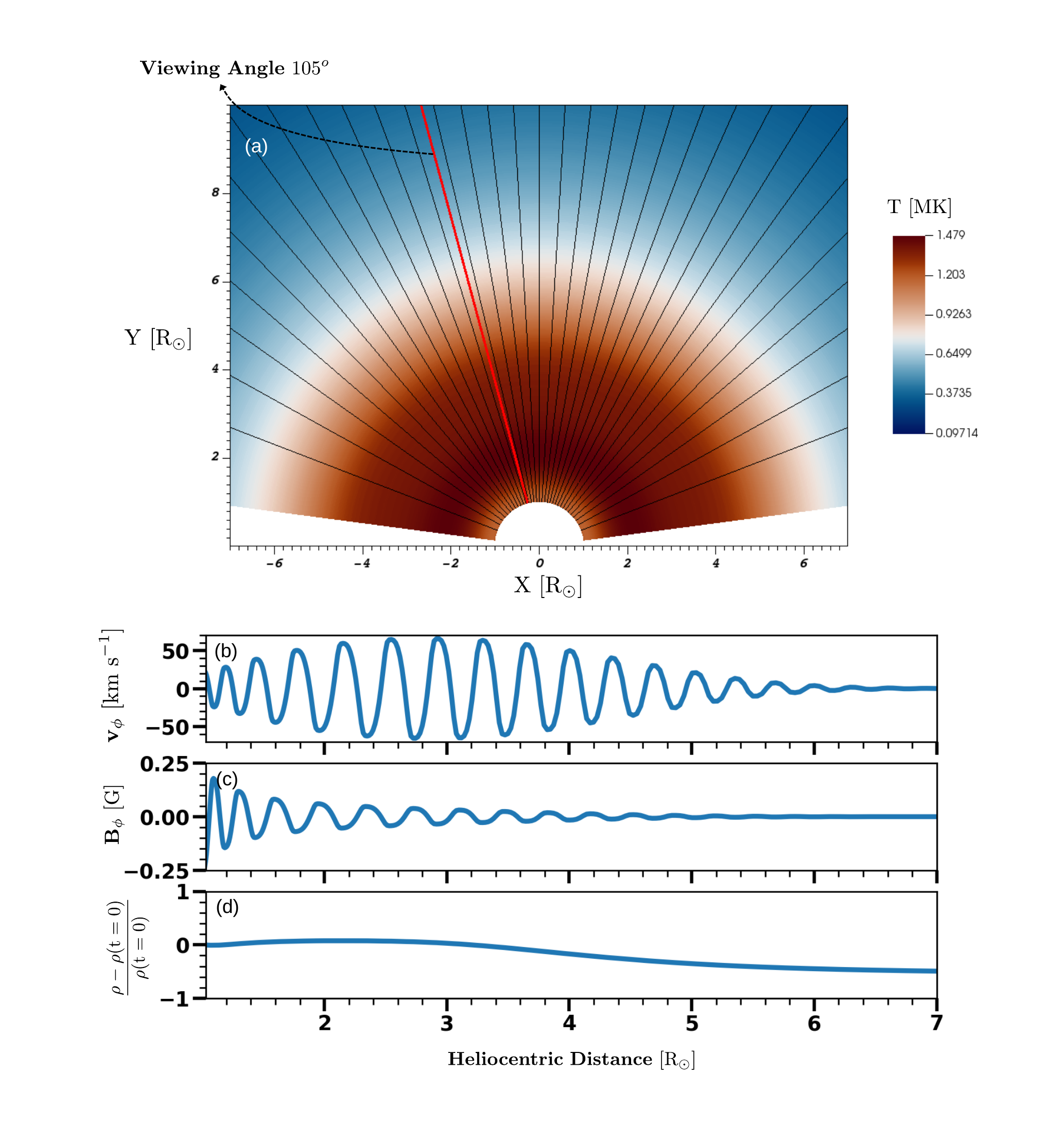}
\caption{Coronal quasi-steady state. Panel (a) shows a snapshot of the plasma temperature upon the injection of a $3~$mHz linearly polarised Alfv\'en wave, with an annotation describing the viewing angle along $105^\circ$. In panels (b) and (c) are shown the out-of-plane $v_\phi$ velocity and $B_\phi$ magnetic field components along the viewing angle. The variations in the density $\rho$ from the quasi-steady values prior to the injection of the Alfv\'en wave are presented in panel (d).}
\label{fig:methods/injAW}
\end{figure}
After achieving a steady-state solar wind by integrating Equations~\ref{eq:mass-cont}-~\ref{eq:def:exp-heating} in time, we introduced Alfv\'{e}nic fluctuations. The Alfv\'{e}n waves were introduced at the coronal base by utilising a time-dependent boundary condition for the Els\"{a}sser variables, defined by
\begin{align}
        \mathbf{z}^\pm_\perp = \mathbf{v}_\perp \pm \frac{\mathbf{B}_\perp}{\sqrt{\mu_0 \rho}}.
        \label{eq:elsasser-var}
\end{align}
We continuously injected the monochromatic and linearly polarized Alfv\'{e}nic fluctuations in the out-of-plane $\phi$ direction by specifying the anti-sunward (outgoing) Els\"{a}sser variable as $\delta\mathbf{z}^- = \mathrm{Z_0}\sin{\left(2\pi f_0 t\right)}~\hat{\mathbf{\phi}}$ at the lower boundary with $\mathrm{Z_0} = 32\sqrt{2}~\mathrm{km s^{-1}}$ being the amplitude and $f_0$ the frequency of the wave. 

In Figure~\ref{fig:methods/injAW} we present the quasi-steady solar wind after the injection of a $3~\mathrm{mHz}$ Alfv\'en wave. In general, the solar wind response to the injected fluctuations depends on the polarization of the waves~\citep{goldstein1978instability, hollweg1971density}. The propagation of the linearly polarised injected Alfv\'en wave causes a fluctuating magnetic field strength which results in the steepening of the Alfv\'en waves themselves~\citep{cohen1974nonlinear}, in addition to generating density fluctuations due to the ponderomotive force~\citep{nakariakov1997alfven}. Due to this, we observed an increase in temperature from 1.2 MK at the lower boundary to 1.4 MK near 3 $R_\odot$, before decreasing again (see Figure~\ref{fig:methods/injAW}(a)). The generation of density fluctuations is a second-order non-linear effect, and is absent in incompressible MHD. In this simulation the density fluctuations are absent as the chosen grid resolution causes the Alfv\'en waves to be damped due to numerical diffusion before the density fluctuations can be generated. This damping ensures that we have only a pure monochromatic Alfv\'en wave in the simulation that has not yet experienced any reflections from large-scale density gradients in the solar wind~\citep{verdini2007alfven, van2011heating}, and confines the waves to be present below $\approx 10~\mathrm{R_\odot}$. Thus, in this study we confined our analysis to the wind below $\approx 10~\mathrm{R_\odot}$.

The Alfv\'enicity, steepening, and absence of density fluctuations in the simulation are illustrated by considering the radial propagation of the injected waves along a viewing angle (annotated in Figure~\ref{fig:methods/injAW}(a)). In Figure~\ref{fig:methods/injAW}, panels (b) and~(c), we present the out-of-plane velocity $v_\phi$ and magnetic field $B_\phi$ components along this viewing angle. Upon comparing the two panels, we observe an anti-correlation between $v_\phi$ and $B_\phi$, which confirms both the Alfv\'enicity and anti-sunward direction of the injected wave. Furthermore, to verify the lack of accompanying density perturbations, we plot in Figure~\ref{fig:methods/injAW}(d) the fluctuating component of the mass density $\Delta\rho/\rho = (\rho-\rho(t=0))/\rho(t=0),$ where $\rho(t=0)$ is the mass density in the coronal volume prior to the Alfv\'en wave injection. The panel shows a large-scale variation in the density, but the absence of any smaller-scale fluctuations.

\subsection{Introducing a coronal mass ejection} \label{subsec:methods/CMEs}
In this study, we do not model the initiation and subsequent eruption of the CME, but instead directly instantiate an erupting plasma structure mimicking an eruptive CME. We superimpose an appropriate plasma structure on the quasi-steady solar wind containing the Alfv\'enic fluctuations to achieve this.

The magnetic field of the CME is modelled as a force-free FR using the Soloviev solution of the Grad-Shafranov (GS) equation~\citep{solov1968theory}. The solutions to the GS equation represent axisymmetric MHD equilibria of magnetized plasmas without flows such that the equilibrium condition 
\begin{align}
    \mathbf{J}\times \mathbf{B} = \nabla P
    \label{eq:GS}
\end{align}
is satisfied where 
$\mathbf{J}$ is the current density given by $\mathbf{J} = \nabla\times\mathbf{B}/\mu_0$, and $P$ is the thermal pressure of the plasma. Once the magnetic structure of the CME is modelled using Equation~\ref{eq:GS} under the assumption of zero-beta ($P=0$) conditions we then populate it with plasma to model a high-density ejecta. The density inside the structure is specified as
\begin{align}
    \rho_\mathrm{cme} =  \frac{\rho_\mathrm{cme, 0}}{2}\left[1 - \cos\left(\pi \frac{d_\mathrm{cme} - d}{d_\mathrm{cme}}\right)\right],
    \label{eq:eruptive-profile}
\end{align}
where $d$ is the distance from the centre of the structure and $d_\mathrm{cme}$ is the radial extent,
and $\rho_\mathrm{cme, 0}$ is the density specified at the centre. This formulation of $\rho_\mathrm{cme}$ ensures a continuous transition from the high-density $\rho_\mathrm{cme, 0}$ CME core to the background density at the edge of the structure. We also initialise the plasma with a constant temperature of $0.5\times 10^6~\mathrm{K}$, and an ejection velocity $\mathbf{v_\mathrm{ej}}$ along the radial direction inside the CME. The constructed CME (Equations~\ref{eq:GS} and~\ref{eq:eruptive-profile}) is then superimposed on the quasi-steady solar wind including the Alfv\'enic fluctuations described in Section~\ref{subsec:methods/AW}. We note that due to the ad hoc specification of the thermal pressure inside the CME and superposition of the structure on the quasi-steady wind the plasma in and immediately surrounding the CME is not in equilibrium causing the FR to expand and propagate.

\begin{figure}[ht]
\centering
\includegraphics[width=0.5\textwidth]{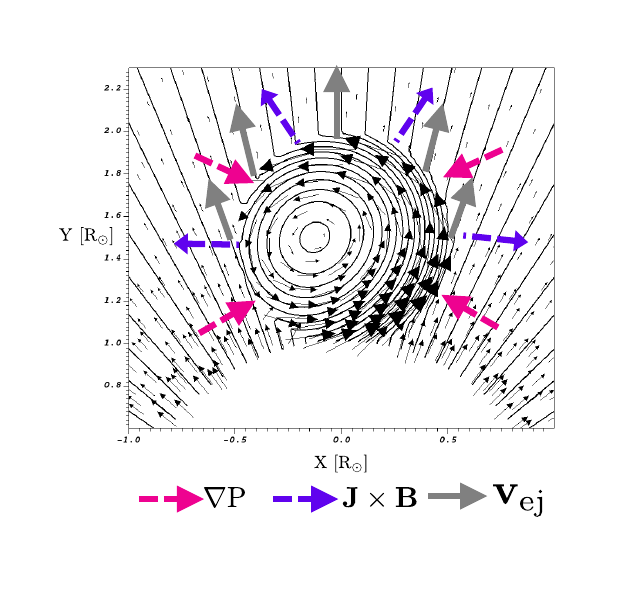}
\caption{CME insertion into the solar wind. The black curves 
show magnetic field lines with the black arrows indicating the direction and relative strength. The figure is annotated with the directions of the $\nabla P$ and $\mathbf{J\times B}$ forces that comprise the Grad-Shafranov equilibrium condition. An initial ejection velocity $\mathbf{v}_\mathrm{ej}$ is given to the CME along the radial direction.}
\label{fig:mc-explained}
\end{figure}

In Figure~\ref{fig:mc-explained} we present a schematic showing the magnetic field configuration and dynamic contributions acting on the CME at the onset. The poloidal field of the FRs we used for this study is oriented in the anti-clockwise direction, as seen by the direction of magnetic field vectors around the FR. This causes them to deflect when reconnecting with the radially outward magnetic field lines. The ejection velocity is also directed in the radial direction.

\begin{figure}[ht]
\centering
\includegraphics[width=0.5\textwidth]{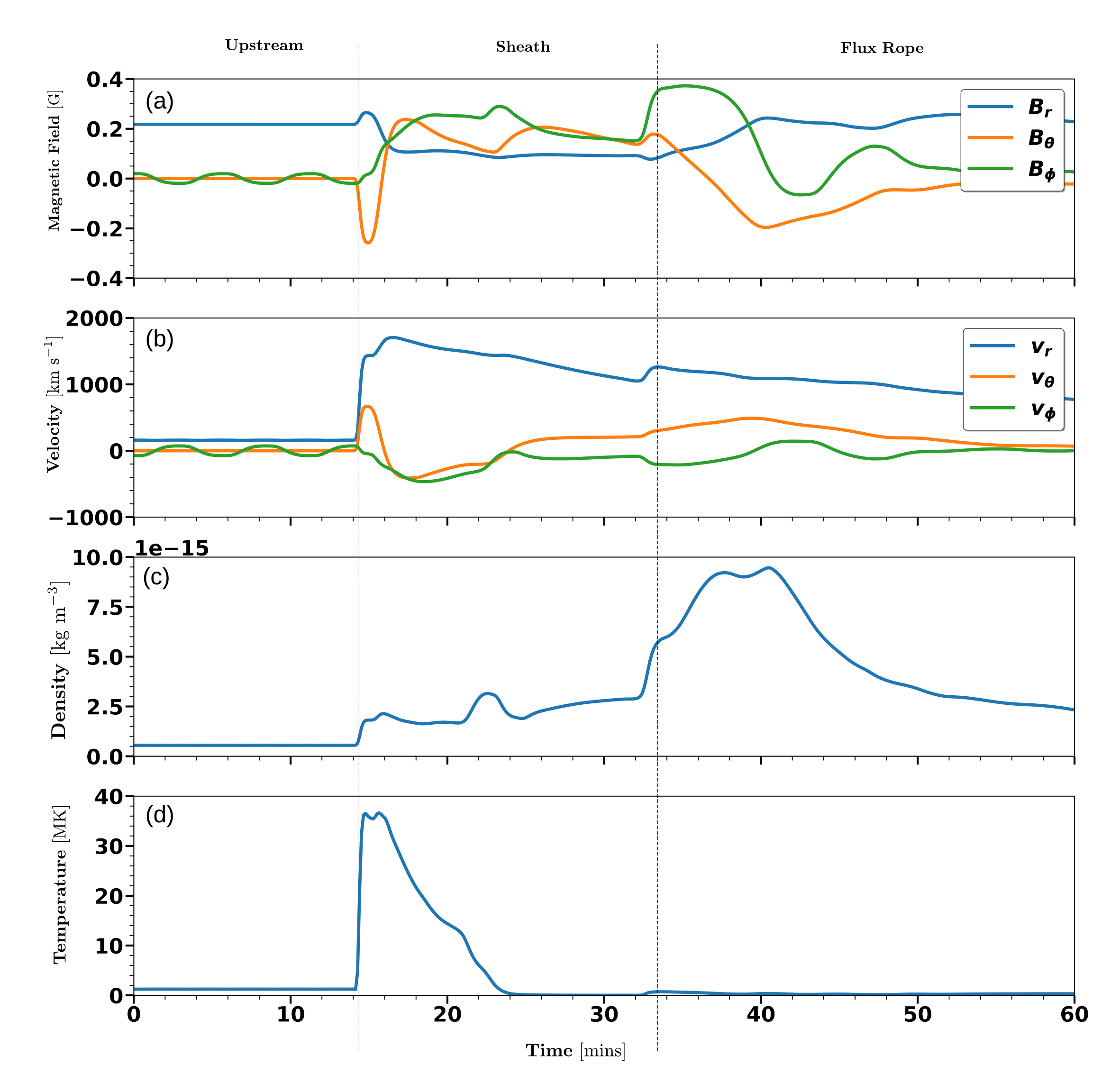}
\caption{CME encounter with a virtual spacecraft located at 5 $R_\odot$ and a viewing angle of $105^\circ$. The vertical lines differentiate the upstream, sheath, and flux rope encountered by the spacecraft.}
\label{fig:virtual-craft}
\end{figure}

The plasma signatures encountered by a virtual spacecraft upon traversing such an FR are shown in Figure~\ref{fig:virtual-craft}. The dashed vertical lines demarcate the upstream, CME sheath, and FR regions. The spacecraft is placed at 5 $R_\odot$, and a viewing angle of 105 degrees with the time axis referenced from the time of CME injection in the simulation. The CME is modelled using an initial speed $v_\mathrm{ej} = 500~$\textrm{km~s}$^{-1}$, peak density of $2\rho_0$ (where $\rho_0$ is the constant mass density at the coronal base at $r=r_0$), and $B_\phi \approx 12$~\textrm{G}. Prior to encountering the CME, the virtual spacecraft measures the pristine upstream solar wind conditions as seen in Figures~\ref{fig:virtual-craft}(a)-(d). We observe anti-sunward Alfv\'enic fluctuations by the anti-correlated variations in $B_\phi$ and $v_\phi$. The first CME-related signature registered is the shock at $t\approx 15$~\textrm{min}. The shock is followed by the CME sheath. In this sheath region we observe larger non-radial flows (compared to the upstream fluctuations) as non-zero values for $v_\theta$ and $v_\phi$ (Figure~\ref{fig:virtual-craft}(b)). The CME sheath is also characterised by an enhanced density and temperature (Figures~\ref{fig:virtual-craft}(c),~(d)) due to the shock transition and plasma piling ahead of the CME. Finally, the spacecraft encounters the FR. It features a smooth variation in $B_\theta$ (Figure~\ref{fig:virtual-craft}(a)) indicating rotation of the field as it crosses the magnetic island initialised in Equation~\ref{eq:GS}. The FR region also has a relatively high density (Figure~\ref{fig:virtual-craft}(c)).

\section{Results}     \label{sec:results}
\begin{figure*}
\centering
\includegraphics[width=\textwidth]{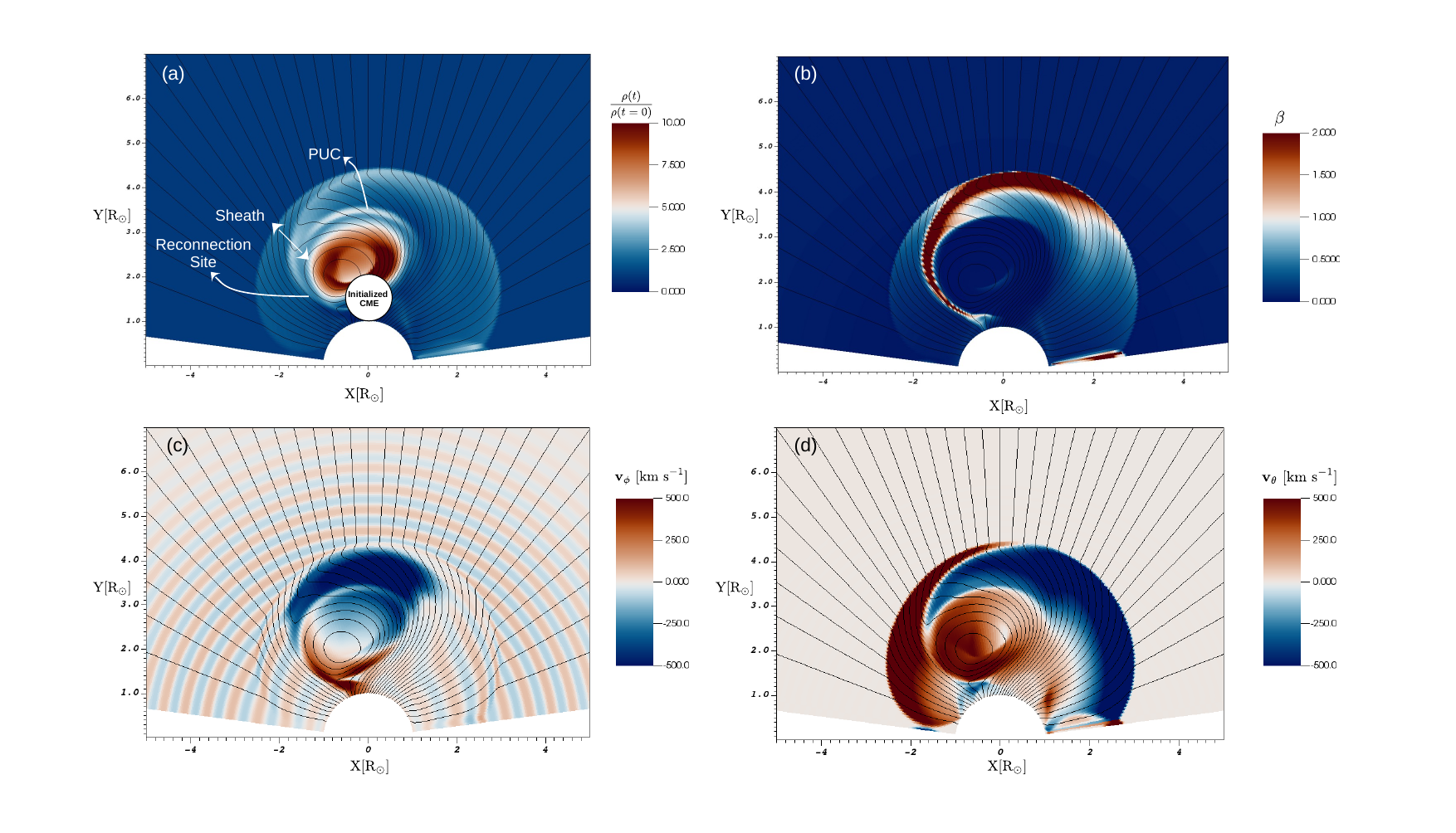}
\caption{Snapshots of CME propagation. The figure presents snapshots of the simulation as the CME is propagating in the low corona at $t = 10.8$ min. In panel (a) the colour intensity denotes the density compression compared to the quasi-steady solar wind, with annotations indicating the PUC, sheath, and reconnection site. The plot in panel (b) shows the plasma beta, and panels (c) and (d) present the out-of-plane velocity component $v_\phi$ and the co-latitudinal (meridional) component $v_\theta$.}
\label{fig:MC}
\end{figure*}

In this section we describe the interaction of the Alfv\'enic fluctuations in the quasi-steady solar wind (Section~\ref{subsec:methods/AW}) with a CME modelled as in Section~\ref{subsec:methods/CMEs} with $v_\mathrm{ej} = 500~$\textrm{km~s}$^{-1}$, peak density of $2\rho_0$, and $B_\phi \approx 12$~\textrm{G}. The CME is deflected in the -X direction as it reconnects with the anti-sunward-directed radial magnetic field line due to the chosen poloidal field direction of the FR. In Figure~\ref{fig:MC}(a)--(c) we show the density compression ratio computed as $\rho(t)/\rho(t=0)$~\citep{pomoell2015DC}, the plasma beta $\beta = p_\mathrm{thermal}/p_\mathrm{magnetic}$, and the out-of-plane (longitudinal) velocity component $v_\phi$ at simulation time $t =$ 10.8 min. 

The initial velocity of the FR ($\mathbf{v_\mathrm{ej}}$) and the out-of-equilibrium $\mathbf{J\times B}-\nabla P$ force allows the plasma of the CME to expand at a rate much higher than the ambient solar wind velocity. This results in the FR driving a fast mode shock. In an ideal MHD system, the maximum density compression ratio at a  shock front is $\frac{\gamma + 1}{\gamma - 1}$~ \citep[e.g.][]{koskinen2011physics}, which in our case, with $\gamma = 5/3$, gives a theoretical maximum compression of $4$. At the shock front, located approximately at $2.5$~$R_\odot$, there is an observed density compression jump from $1$ in the upstream region to $\approx 2$ inside the sheath at the flank of the CME, and to $\approx 3$ near the head-on region of the CME. The FR is driving a shock as a result of the large difference between the CME ejection velocity and upstream solar wind velocity. The 
FR trails behind the leading shock front and is identified by the closed field lines forming the magnetic island. The sheath is the region between the shock and FR. 
CME sheath regions are often characterised by non-radial flows and build-up of density in a pile-up compression region (PUC) ~\citep{das2011evolution}. In our simulation the presence of non-radial flows is due to the draping of the flow around the magnetic island in the sheath~\citep{siscoe2008ways}. This draping causes the formation of an oblique shock, which in turn causes large-scale flows to be generated to maintain the non-radial continuities in the Rankine-Hugoniot jump conditions. Additionally, the compression of plasma in the sheath region causes the formation of a PUC. Figure~\ref{fig:MC}(a) is annotated with markings denoting the PUC, the sheath region, and the location of the reconnection site causing the CME to deflect. The reconnection at the CME flank reduces magnetic flux at this location, while the field still drapes around the FR in the opposite flank. This drives a strong magnetic field gradient that deflects the CME.

In Figure~\ref{fig:MC}(b) we plot the plasma beta in the simulation to investigate whether plasma dynamics are dominated by the magnetic field (low $\beta$) or gas dynamics (high $\beta$). We see that the whole steady-state solar wind upstream of the shock has $\beta \ll 1$, which indicates a frozen-in plasma condition is strongly met. We observe a region of high $\beta$ inside the sheath as we view the CME head-on and around the reconnection site. A comparison with Figure~\ref{fig:MC}(a) shows that this high-$\beta$ region occurs when we encounter a region of enhanced density inside the sheath. The FR is isolated from the surrounding sheath region and maintains a low $\beta$.

Finally, in Figures~\ref{fig:MC}(c) and \ref{fig:MC}(d) we present the $v_\phi$ and $v_\theta$ components in the simulation, respectively. At the CME flanks, we see that the solar wind perturbations in $v_\phi$ are modified by the shock. The radially directed wave vectors upstream of the shock are modified downstream to reflect the non-radial topology of the shock front. However, in the regions of enhanced $\beta$ from Figure~\ref{fig:MC}(b), there are significant flows in the $\pm\phi$ directions, as shown by the large $v_\phi$ magnitudes. These are the non-radial flows that are generated as a result of the structure of the FR that causes the solar magnetic field to drape around it, as well as the strong guide field of the FR affecting the flow of the surrounding plasma. The spatial extent of the non-radial flows, as depicted by the dark blue region in Figure~\ref{fig:MC}(c), indicates a similarity in size to the wavelength of Alfv\'en waves at the flanks of the CME. The $v_\theta$ component does not have any perturbations upstream as expected. However, downstream of the shock, we see large flows in $\theta$ as the FR sweeps away the surrounding plasma as it propagates. 

\subsection{CME modified solar wind fluctuations}
\begin{figure}[ht]
        \centering
        \includegraphics[width=0.5\textwidth]{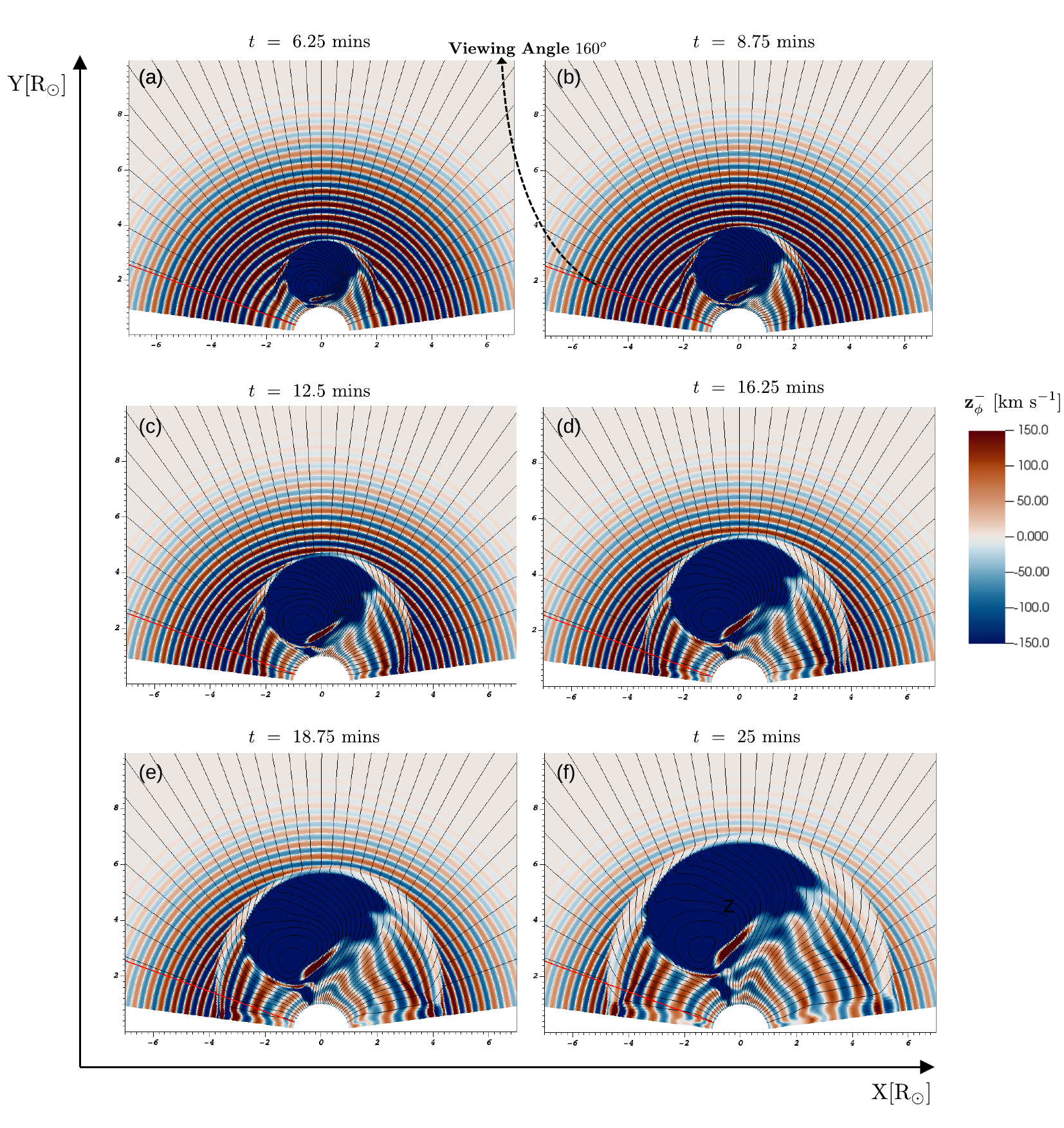}
        \caption{Snapshots of Els\"asser variables. The figure presents the anti-sunward-propagating Els\"asser variable $z_\phi^- = v_\phi - B_\phi/\sqrt{\mu_0\rho}$ during the CME evolution shown at various times. The figures are annotated with a viewing angle of $160^\circ$ corresponding to the CME flank.}
        \label{fig:evolution}
\end{figure}

The cut through the CME flanks in Figure~\ref{fig:MC}(c) shows that the frequency of the upstream solar wind fluctuations decreases downstream of the shock. Alfv\'{e}nic fluctuations such as these are characterised by a correlation in the velocity and magnetic field~\citep{belcher1969large} and can be identified by the accompanying Els\"{a}sser variables (Equation~\ref{eq:elsasser-var}). Fluctuations can be identified by subtracting the mean plasma flow speed from the accompanying Els\"{a}sser variable. The simulation snapshots of the anti-sunward-propagating Els\"asser variable $z_\phi^-$ at various times is shown in Figure~\ref{fig:evolution}. This figure is annotated with the viewing angle of $160^\circ$ corresponding to the flank of the CME.

The panels (a) and (b) present the anti-sunward Els\"asser variable 6.25 and 8.75 minutes after the onset of the eruptive event, respectively. The significant negative value of $z_\phi^-$ in the figure is due to the positive $B_\phi$ inside the FR. The large positive $B_\phi$ field compresses the plasma ahead of it, causing the large negative valued $z_\phi^-$. This positive $B_\phi$ along with the anti-clockwise direction of the poloidal field around the FR (Figure~\ref{fig:mc-explained}) denotes a positive (right-handed) chirality for the FR. 
At the flanks, the initially expanding CME amplifies the imposed fluctuations on the flanks as it `drags' the solar wind at speeds higher than the ambient Alfv\'en velocity prior to shock formation (Figures~\ref{fig:evolution}(a)-(d)). After the formation of a shock along the $160^\circ$ viewing angle in panel (e), these CME-modified anti-sunward fluctuations are also present in the downstream region.

\subsubsection{Shock transmitted solar wind fluctuations}
\label{subsec:shock-transmission}
\begin{figure}[ht]
\centering
\includegraphics[width=0.5\textwidth]{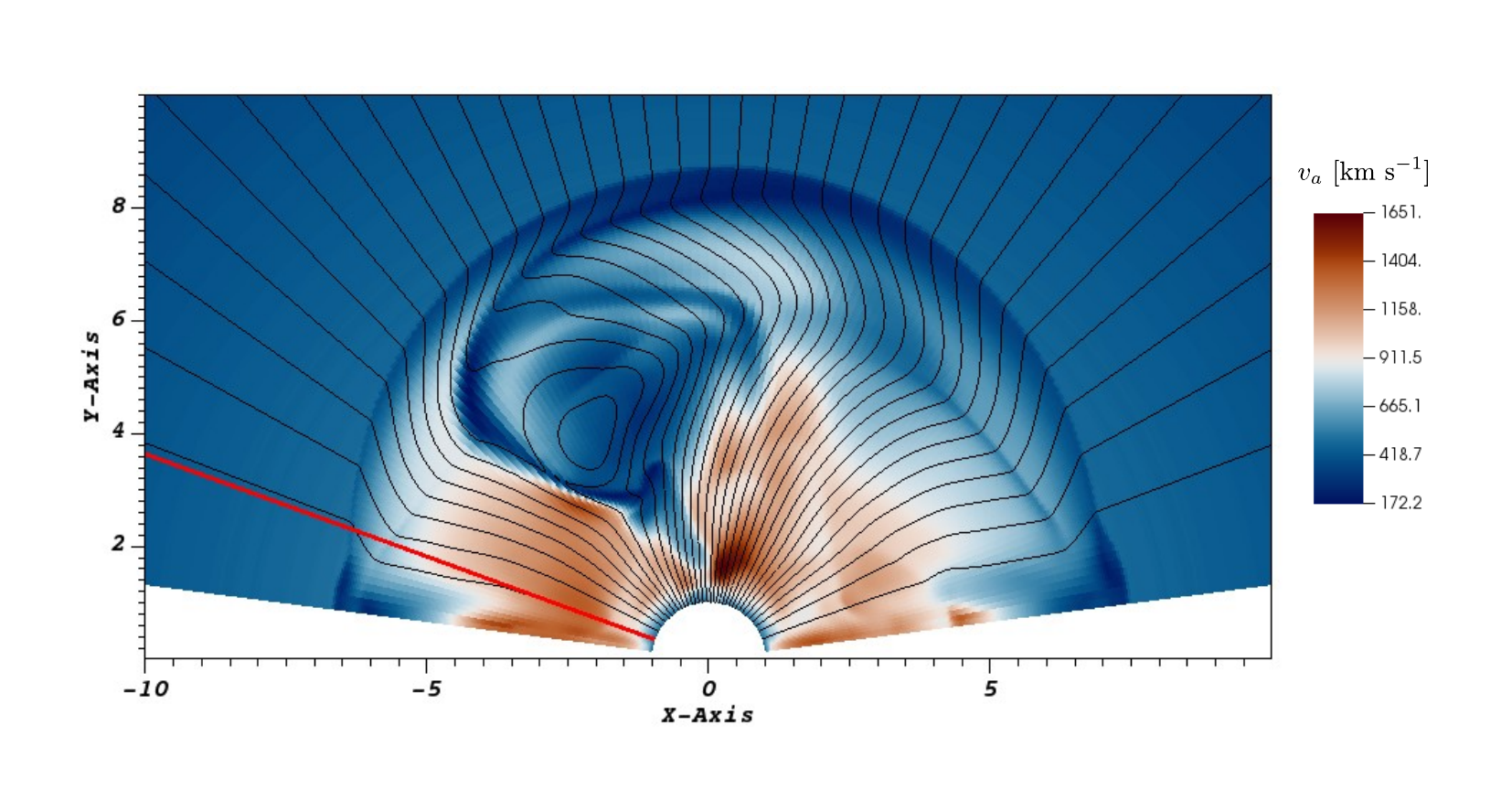}
\caption{Snapshot of the Alfv\'en velocity, defined as $v_a = \mathbf{v_a}\cdot\hat{\mathbf{b}}$, at $t = 37.5$~\textrm{min}. The figure is annotated to show the CME flank at the viewing angle $160^\circ$.}
\label{fig:aw}
\end{figure}

The presence of a shock modifies the upstream anti-sunward solar wind fluctuations as they are transmitted (propagating anti-sunward) and reflected (propagating sunward) downstream of the shock~\cite[e.g.][]{vainio1998alfven, vainio1999self}. If a medium is stationary, a wave propagates conserving its frequency. In the shock frame, the fluid structure is quasi-stationary on the timescale it takes for the wave to be transported through the shock, so Alfv\'en waves conserve their frequency in the shock frame. Another boundary condition at the shock for the wave vector comes from the conservation of the tangential wave length. Thus, for a transmitted, outward-propagating Alfv\'en wave,
\begin{eqnarray}
        k_{1,n}(u_{1,n}-v_{a1,n}) &=& k_{2,n}(u_{2,n}-v_{a2,n}),
        \label{eq:conservation-freq}\\
    k_{1,t} & = &k_{2,t}
,\end{eqnarray}
where $\mathbf{u}$ is the fluid velocity in the shock frame, $\mathbf{k}$ is the wave vector, $\mathbf{v}_a$ is the Alfv\'en velocity, the subscripts 1 and 2 denote the upstream and downstream regions, and the subscipts $n$ and $t$ denote the normal and tangential components of the vector quantities in relation to the shock surface normal. As the normal component of the magnetic field is conserved at the shock, $v_{a2,n} = v_{a1,n}/\sqrt{X}$, where $X = \rho_2/\rho_1 = u_{1,n}/u_{2,n}$ is the compression ratio of the shock. Thus, the downstream wave number
\begin{equation}
    k_{2,n} = k_{1,n}X\frac{M_A-1}{M_A-\sqrt{X}},
\label{eq:anti-sunward-kvec}
\end{equation}
where $M_A=u_{1,n}/v_{a1,n}$ is the Alfv\'enic Mach number, showing that the wavelength in the shock normal direction is compressed by a factor exceeding the gas compression ratio of the shock. For a low-Mach-number ($M_A\lesssim 2$) quasi-parallel fast-mode shock propagating in a low-$\beta$ plasma, the compression ratio is approximately $X\lesssim M_A^2$ \citep{vainio1999self}, implying that the wave compression can be very significant. We note that at the limit of a switch-on shock ($X=M_A^2$), wave compression becomes infinite. For a reflected wave (i.e. the case where the downstream wave is propagating towards the Sun), the wave compression is less significant,
\begin{equation}
    k_{2,n} = k_{1,n}X\frac{M_A-1}{M_A+\sqrt{X}},
\label{eq:sunward-kvec}
\end{equation}
in particular for a low-Mach-number shock.
Thus, we expect the upstream Alfv\'en wave to significantly decrease in wavelength as it propagates downstream of the CME shock. Therefore, the expected composition of the downstream anti-sunward solar wind fluctuations as a consequence of the shock transmission consists of a long-wavelength component as the CME flank drags the waves that were transmitted downstream through the early quasi-perpendicular phase of the shock on a given field line ($k_t$ is conserved) and a short-wavelength component due to the Alfv\'en wave transmission across the quasi-parallel part of the shock. The waves transmitted in the early quasi-perpendicular stage (Figure~\ref{fig:evolution}) have a longer wavelength as they experience a higher Alfv\'en velocity downstream of the CME shock (Figure~\ref{fig:aw}) causing the increase in wavelength.

\subsubsection{Fluctuations around the CME shock}
\label{subsec:fluctuations-around-shock}
\begin{figure}[ht]
        \centering
        \includegraphics[width=0.5\textwidth]{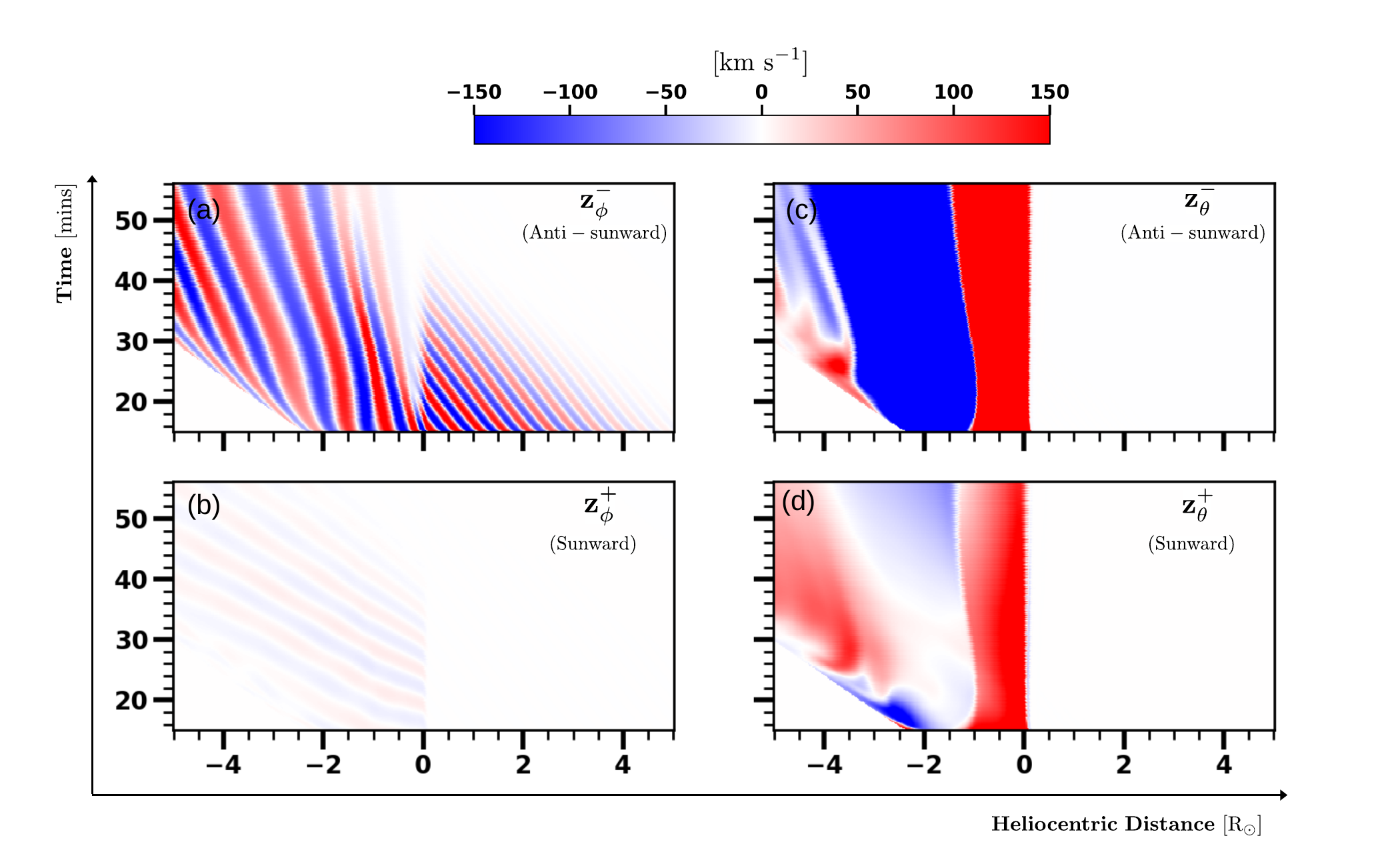}
        \caption{Evolution of the Els\"asser variables at the CME flank. The spatio-temporal evolution of the Els\"{a}sser variables for the non-radial directions in the frame of reference of the shock ($x = 0$) is presented. The quantities are shown for a viewing angle of $160^\circ$, and an Alfv\'enic fluctuation frequency of $3$~\textrm{mHz}. The x-axis denotes the shock neighbourhood in units of $R_\odot$;  positive values indicate the solar wind and negative values indicate the region downstream of the shock.}
        \label{fig:results/elss-3mHz}
\end{figure}

In Figure~\ref{fig:virtual-craft}, we observe the presence of a variety of radial (via the shock propagation) and non-radial (in the sheath plasma) enhancements in the velocity. In Figure~\ref{fig:results/elss-3mHz} we attempt to exclude the radial flow by viewing the Els\"{a}sser variables in the frame of reference of the shock. The CME shock is detected by locating a jump in density compression that exceeds a factor of 2 and is always placed at the $x=0$ coordinate with the Els\"{a}sser variables shown in the neighbourhood of 5 $R_\odot$. The positive x-axis values are upstream of the shock, and the negative x-axis are downstream.

In Figure~\ref{fig:results/elss-3mHz} we present the sunward ($z_{\phi, \theta}^+$) and anti-sunward Els\"{a}sser variables for a viewing angle of $160^\circ$ (CME flank) and the Alfv\'{e}n wave with frequency 3~mHz injected in the quiet solar wind. The x-axis represents the distance along the given viewing angle, and the y-axis is the simulation time. The magnitude of the Els\"{a}sser variables are described using the colour intensity. In Figure~\ref{fig:results/elss-3mHz}(a) the upstream solar wind fluctuations can be observed as they are incident onto the CME shock from $x > 0$. The upstream Alfv\'en waves in the simulation have significant amplitudes until $\approx 10~\mathrm{R_\odot}$ and therefore disappear beyond $\mathrm{t} \approx 40~\mathrm{min}$. Further into the downstream $x < 0$ we see the long-wavelength component as the CME propagation modifies the downstream waves. The white region between these two regions is where the short-wavelength component of the shock-transmitted waves should be present, as expected based on the analysis in Section \ref{subsec:shock-transmission}. This white region corresponds to a similar region in Figures~\ref{fig:evolution}(e)-(f) around the CME shock where the upstream waves should be compressed. Additionally, we see the generation of sunward fluctuations in Figure~\ref{fig:results/elss-3mHz}(b) due to the interaction of solar wind fluctuations with the CME shock (Equation~\ref{eq:sunward-kvec}). In the $\theta$ direction, we do not see any fluctuations upstream, as expected in Figure~\ref{fig:MC}(d). However, in the shock neighbourhood we can see the effect of non-radial flows due to it being a non-radial shock, which was also seen in Figure~\ref{fig:results/elss-3mHz}(d).

\begin{figure}[ht]
        \centering
        \includegraphics[width=0.5\textwidth]{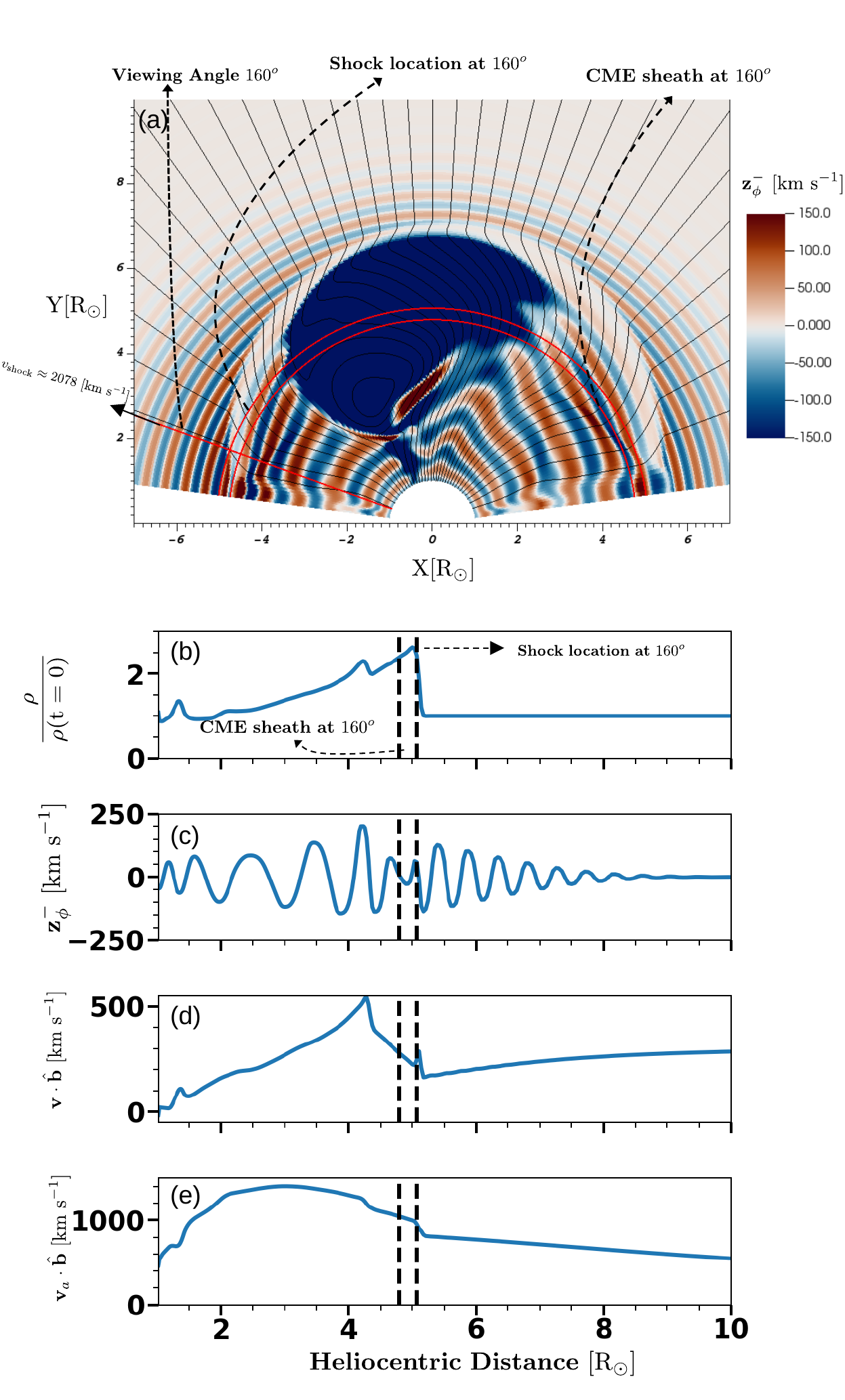}
        \caption{Shock at the CME flank. Panel (a) is a simulation snapshot at $t=25~$min of the anti-sunward Els\"asser variable $z_\phi^-$ with annotations describing the viewing angle along $160^\circ$, the shock location, the approximate beginning of the CME sheath, and the approximate shock velocity $v_\mathrm{shock}$. Panels (b) and (c) are the density compression and $z_\phi^-$ along the viewing angle, respectively. Panels (d) and (e) present the fluid velocity and Alfv\'en speed along the direction of the background magnetic field.}
        \label{fig:compression}
\end{figure}

The white region in Figure~\ref{fig:results/elss-3mHz}(a) is further investigated through Figure~\ref{fig:compression} where the density compression, the anti-sunward Els\"asser variable, the flow speed ($\mathbf{v\cdot b}$), and the Alfv\'en speed along the background field ($\mathbf{v}_a\cdot\mathbf{b}$) are presented at $t = 25~$min. Panel (a) of the figure is an annotation of Figure~\ref{fig:evolution}(f) with the location of the CME shock and the approximate beginning of the sheath, where we start observing the long-wavelength fluctuations. The  white  region thus corresponds to the location between these two markers. Panel (b) shows the density compression utilised in identifying the shock, and panel (c) is the anti-sunward Els\"asser variable. The average shock velocity at the $160^\circ$ viewing angle between $t = 6.25~$min to $t = 31.25~$min is found to be $\approx 2078~$km s$^{-1}$, as annotated in panel (a). The shock is associated with a gas compression ratio of $\approx 2$ (panel (d)), with the Alfv\'en speed increasing from upstream to downstream (panel (e)). Then, downstream of the shock (through Equation~\ref{eq:sunward-kvec}) the wavelength of the upstream wave would be compressed by about three times. The absence of the anticipated compression of the upstream Alfv\'en wave in our simulation indicates that the spatial grid does not adequately resolve this specific region. This causes the transmitted waves to be of lower amplitude in this location, as observed in panel (c), signifying numerical dissipation. Therefore, the downstream fluctuations plotted in Figure~\ref{fig:results/elss-3mHz} do not contain the additional shock-compressed Alfv\'en waves. However, the restricted grid resolution for this simulation is necessary to sustain a monochromatic Alfv\'en wave before the CME injection by numerically damping the waves before their decay. In Appendix~\ref{appendix:shock-compression-aw} we present a modified simulation where the shock-compressed waves are captured. The results in the Appendix show an adherence to the expected composition of the downstream waves (see Section~\ref{subsec:fluctuations-around-shock}), with the downstream waves becoming modified compared to Figure~\ref{fig:results/elss-3mHz} only after $\mathrm{t} \approx 20~\mathrm{min,}$ due to possible wave steepening. Thus, in practice the downstream solar wind fluctuations would contain long- and short-wavelength components due to the CME passage and upstream waves transmission, respectively, prior to the development of further non-linear interactions.

\begin{figure}[ht]
        \centering
        \includegraphics[width=0.5\textwidth]{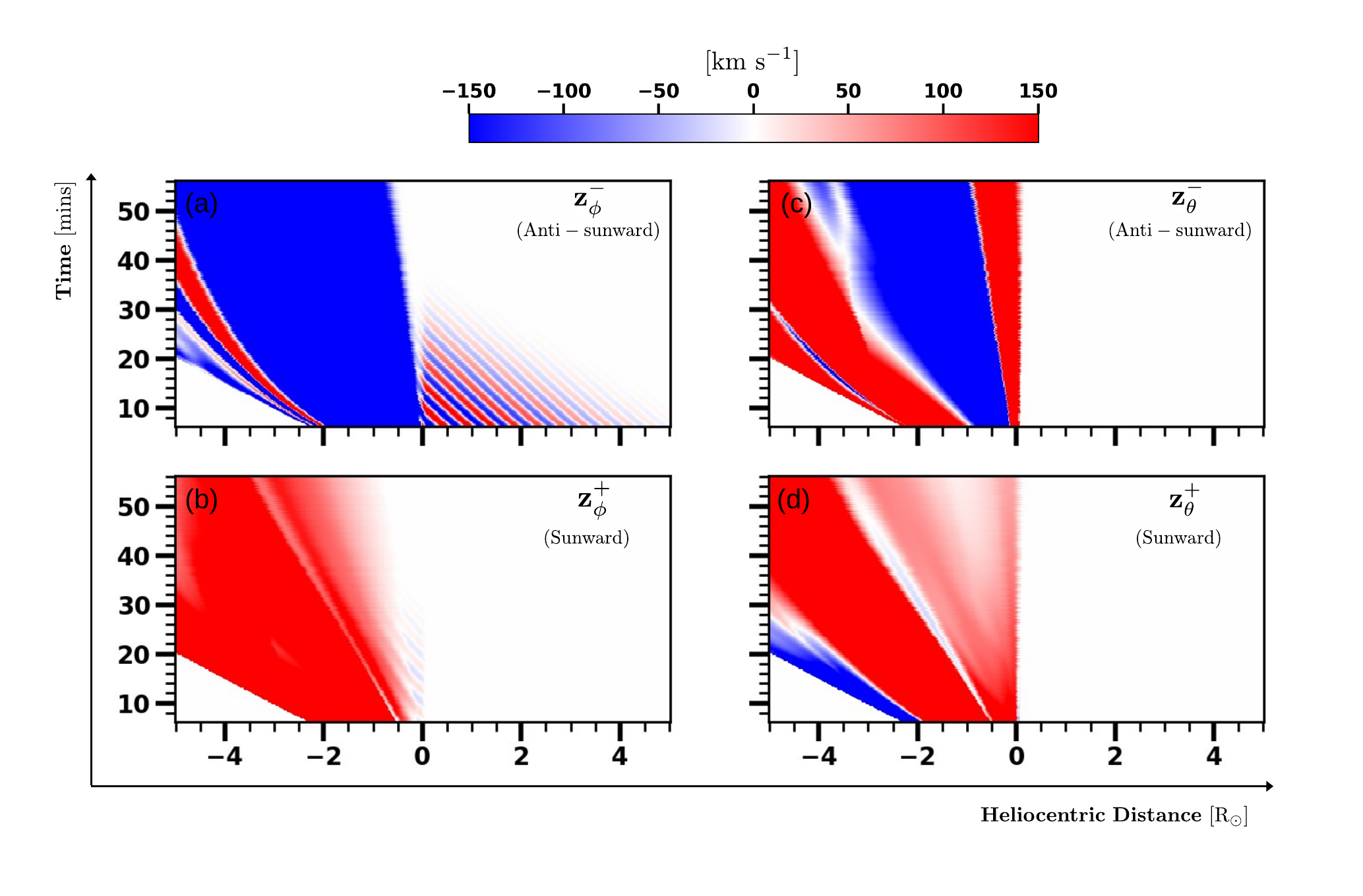}
        \caption{Evolution of the Els\"asser variables at the CME nose, similar to Figure~\ref{fig:results/elss-3mHz}, but with a viewing angle of $105^\circ$. The x-axis similarly denotes the shock neighbourhood in units of $R_\odot$ with the shock centred at $x=0$.}
        \label{fig:results/elss-3mHz-105}
\end{figure}

In Figure~\ref{fig:results/elss-3mHz-105}, we show the solar wind fluctuations around the shock when viewing the CME head-on instead of the flank. Panel (a) shows the upstream solar wind fluctuations incident onto the shock. A similar white region corresponds to the region where the incident waves would be compressed. However, in the far downstream region $x < 0$, we only observe large non-radial flows as the large positive guide field $B_\phi$ of the FR affects the surrounding plasma to generate a non-radial flow. These non-radial flows are additionally observed in the sunward component (panel (b)). In the $\theta$ directions (panels (c) and (d)) the large flows are generated due to the non-radial topology of the CME shock. 

Therefore, a primary difference between solar wind fluctuations downstream of the CME shock for a head-on encounter (Figure~\ref{fig:results/elss-3mHz-105}) compared to a flank encounter (Figure~\ref{fig:results/elss-3mHz}) is the absence of long-wavelength amplified fluctuations, which are comparable in size to the non-radial flows. Through Figure~\ref{fig:compression}(a) it is seen that the CME shock is non-radial as the shock velocity is greater head-on (the direction where the FR is expanding) than at the flank. This indicates that the wavelengths of the shock-compressed upstream waves differ as the compression depends on the Alfv\'en Mach number in the shock frame of reference. Furthermore, as the shock expands faster than the ambient Alfv\'en speed, we expect different characteristics of the fluctuations closer to the shock (containing a mix of shock-transmitted and already present fluctuations) and further downstream (with the CME-amplified fluctuations).

\section{Formation of the CME Sheath}\label{sec:cme-sheath}
In Section~\ref{sec:results} we discussed the dependence of the CME sheath fluctuations on the upstream solar wind conditions and the shock properties. The interaction of the solar wind fluctuations with the CME shock gave rise to both sunward and anit-sunward Alfv\'enic fluctuations at the CME flank (Figures~\ref{fig:results/elss-3mHz}), along with the compression of anti-sunward upstream waves. In addition, the CME sheath contains non-radial flows due to the magnetic structure of the FR and the non-radial CME shock (Figures~\ref{fig:MC}(c)-~\ref{fig:MC}(d)). The extent of the non-radial flows, represented by the dark blue region in Figure~\ref{fig:MC}(c) and Figure~\ref{fig:evolution}, suggests that their spatial extent is comparable to the Alfv\'en waves at the CME flanks. This limitation could hinder the presence of large-amplitude Alfv\'enic fluctuations in the presence of similarly large non-radial flows. Thus, we now investigate the influence of the Alfv\'en waves on the growth of the CME sheath region and propagation of the CME shock. This allows us to understand the effect of Alfv\'en waves on the shock properties and to infer the development of non-radial flows as the CME propagates further in the solar wind.

Previous studies have shown significant variations of the sheath thickness based on the physical properties of the CME, more precisely the properties of the CME  FR and the shock compression ratio~\citep{RUSSELL2002527}. Thus, we investigate how the large-scale structure of the sheath depends on the density and injection velocity of the FR driving it, and the frequency of the Alfv\'{e}nic fluctuations that are present in the solar wind.
These different cases, studied by varying a selection of the parameter values of the simulation set-up, including the case studied in the previous sections (henceforth designated as C1) are detailed in Table~\ref{tab:cme-list}. We find that the large-scale structures of the sheath, such as a PUC, high-speed flows, and magnetic field line draping, are similar for all the cases considered in Table~\ref{tab:cme-list}.

\begin{table}[ht]
\centering
\begin{tabular}{|c|cc|c|}
\hline
\multirow{2}{*}{\textbf{Case}} & \multicolumn{2}{c|}{\textbf{CME Parameters}}                                                                                                                           & \multirow{2}{*}{\textbf{$f_\mathrm{inj}$ {[}mHz{]}}} \\ \cline{2-3}
                               & \multicolumn{1}{c|}{\begin{tabular}[c]{@{}c@{}}$\mathbf{v}_\mathrm{ej}$ [km s$^{-1}$]\end{tabular}} & \textbf{$\rho(t=0)/\rho_0$} &                                                                                \\ \hline
C1                             & \multicolumn{1}{c|}{500}                                                                                         & 1                                                   & 3                                                                              \\ \hline
C2                             & \multicolumn{1}{c|}{500}                                                                                         & 2                                                   & 3                                                                              \\ \hline
C3                             & \multicolumn{1}{c|}{500}                                                                                         & 0.5                                                 & 3                                                                              \\ \hline
C4                             & \multicolumn{1}{c|}{1000}                                                                                        & 1                                                   & 3                                                                              \\ \hline
C5                             & \multicolumn{1}{c|}{500}                                                                                         & 1                                                   & 5                                                                              \\ \hline
C6                             & \multicolumn{1}{c|}{500}                                                                                         & 1                                                   & 0                                                                              \\ \hline

\end{tabular}
\caption{Model parameters used in the different simulation runs. 
The ejection velocity $\mathbf{v}_\mathrm{ej}$ is the initial velocity imparted to the CME along $90^\circ$, the $\rho/\rho_0$ parameter is the density imparted to the CME as a multiple of the low-coronal boundary density, and $f_\mathrm{inj}$ denotes the frequency of the injected  Alfv\'enic perturbations.}
\label{tab:cme-list}
\end{table}


\begin{figure*}[ht]
\centering
\includegraphics[width=\textwidth]{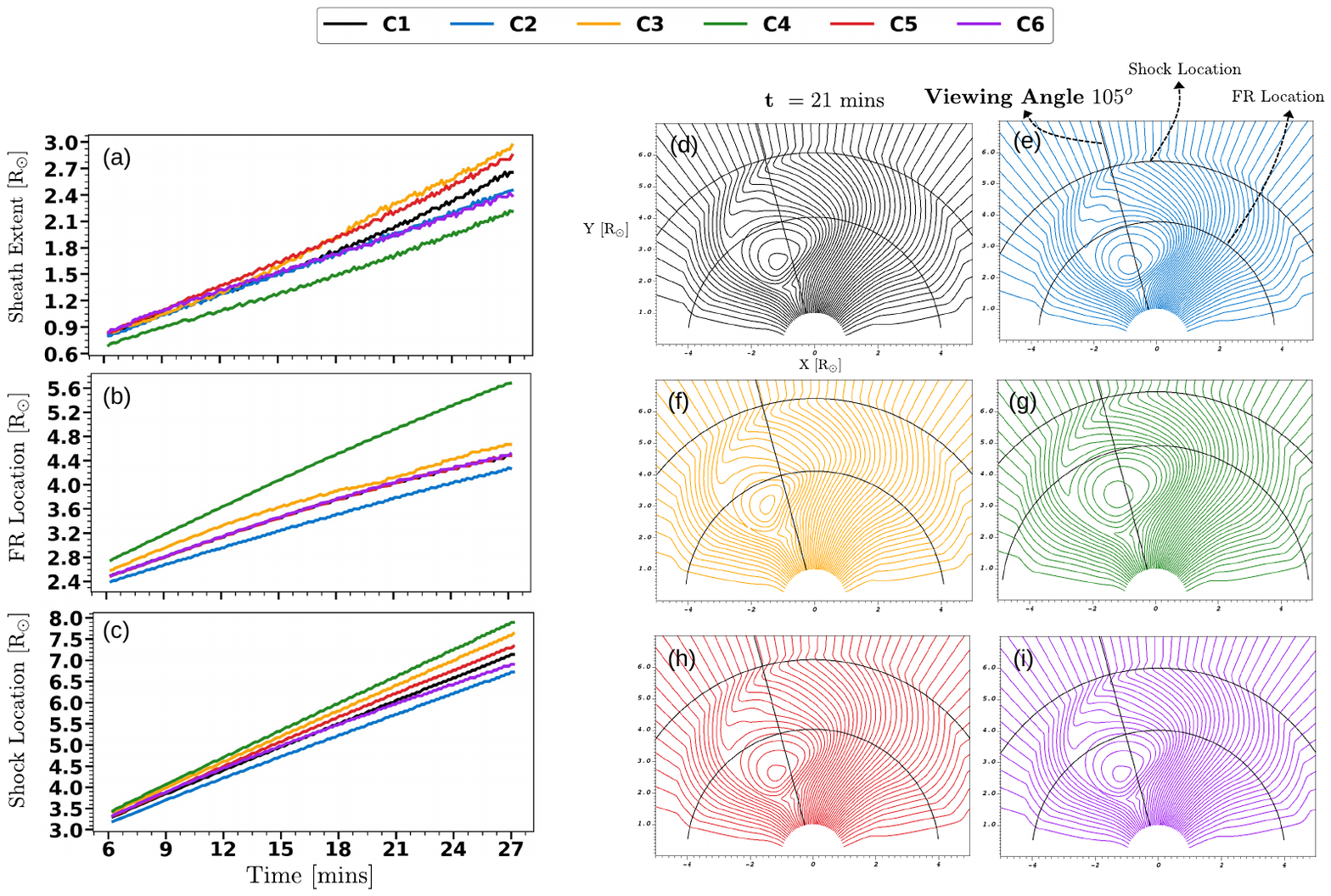}
\caption{Formation of the CME sheath in the different simulation runs. Shown are the evolution of the radial extent of the sheath (a), the flux rope leading edge (b), and the location of the CME shock along a viewing angle of $105^\circ$ (c) for the simulation runs detailed in Table~\ref{tab:cme-list}. The individual runs at time $t = 21$ min are visualised in (d)-(i); the magnetic field lines are colour-coded according to the case number. In each panel (d)-(i), the viewing angle, the flux rope, and shock locations are also indicated.}
\label{fig:results/sheath}
\end{figure*}

To quantify the differences, for each model run 
the extent of the sheath, location of the FR, and the shock location for a viewing angle of $105^\circ$ 
(head-on encounter)
are computed and presented in Figure~\ref{fig:results/sheath}(a)-(c) as a function of time. The computation of the shock's location relies on the density compression ratio, while the positioning of the FR is determined by identifying the first closed magnetic field contour encountered along the viewing angle directed towards the Sun. Subsequently, in Figures~\ref{fig:results/sheath}(d)-(i), we display the snapshot of the simulation for the various cases (Table~\ref{tab:cme-list}) at $t = 21$~\textrm{min} from the event onset. These snapshots are overlaid with markers displaying the viewing angle, shock location, and the FR leading edge location.

Panels (b) and (c) in Figure~\ref{fig:results/sheath} show that the lower the density of the FR, the faster the FR and its leading shock propagate through the corona; this is seen by comparing the high-density (C2; blue curve), nominal density (C1; black curve), and low-density (C3; orange curve) cases.
We note that for these three cases the initial FR speed and the frequency of injected fluctuations were the same ($500$ km s$^{-1}$ and 3 mHz, respectively). From these cases, the low-density FR (C3) that propagates fastest through the solar corona has the widest sheath (panel a). This dependency of the propagation speed on density can be understood by considering 
the deflection of case C3; as explained in Section~\ref{sec:results}, deflection is expected to result from magnetic reconnection at the FR boundaries. The comparison between simulation snapshots in Figures~\ref{fig:results/sheath}(d)-(f) show that the low-density case C3 deflects more than the higher-density cases C1 and C2. This deflection causes our selected viewing angle to focus on the flank of the CME for C3, while for C1 and C2 their higher inertia prevents them from deflecting and they are probed head-on, as intended. 

Next we explore the effect of FR injection velocity. The shock and the FR for C4 (green curve) propagate faster through the corona than for C1, which has the same density but slower speed (see Fig.~\ref{fig:results/sheath}, panels (b) and (c)).
This results in a much smaller sheath thickness for C4 than for C1 (Figure~\ref{fig:results/sheath}(a)). Finally, when we increase the Alfv\'{e}nic fluctuation frequency in C5, and compare it to run C1, which otherwise has identical parameters, there is no notable difference in the FR location (Figure~\ref{fig:results/sheath}(b)), but the shock propagates faster (Figure~\ref{fig:results/sheath}(c)), causing the sheath extent to increase (Figure~\ref{fig:results/sheath}(a)). Moreover, when not injecting any Alfv\'{e}nic fluctuations, as is the case for run C6, we still see no difference in the FR location compared to C1, but the shock propagates more slowly and the sheath extent decreases.

The variations in the shock evolution (Figure~\ref{fig:results/sheath}(c)) for the different cases can be understood through the steepening of the MHD waves. The initial eruption of the CME onto a quasi-steady solar wind generates a fast wave propagating ahead of the CME as it initially strongly expands in the surrounding plasma. If we assume the wave driven by the FR to be a pressure wave, then as this wave propagates it locally compresses the plasma and increases the local sound speed. This would cause the next pressure wave pulse generated by the outward-propagating FR to catch up to the preceding wave modes, thus causing a shock to be generated by the steepening of large-amplitude compressive disturbances. In the general case, the fast wave generated by the FR eruption would be a MHD wave. The rate of steepening a fast mode MHD wave, with no additional assumptions other than the compressibility of the medium, was previously derived~\citep[e.g.][]{kantrowitz1966plasma} to be given by
\begin{align}
    \gamma_s = \omega \frac{\delta \rho}{\rho}\left[1 + \frac{1}{2}\frac{(\gamma-1)v_A^2 c_s^2\sin^2\theta + (v_\mathrm{ph}^2 - c_s^2)^2}{v_A^2 c_s^2\sin^2\theta + (v_\mathrm{ph}^2 + c_s^2)^2}\right].
    \label{eq:steepening-rate}
\end{align}
Here $\omega$ is the wave frequency, $v_A$ the Alfv\'en speed, $c_s$ the sound speed, $v_{ph}$ the phase speed of the wave, and $\theta$ the wave normal angle relative to the magnetic field. The steepening rate depends primarily on the compressibility of the medium ($\delta\rho/\rho$), where $\rho$ corresponds to the undisturbed solar wind density, with minor contributions from the term in brackets~\citep{kennel1985, tsurutani1987steepened}.
Among the CME runs described in Table~\ref{tab:cme-list}, the high-density FR (case C2) corresponds to an increased $\delta \rho/\rho$, while for the low-density FR (case C3) $\delta \rho/\rho$ is smaller than for C1. As a consequence, for C2 the wave steepens to a shock more quickly (at a lower starting height) than for C1, while for C3 the shock forms later. The shock locations in Figure~\ref{fig:results/sheath}(c) grow linearly with time, which indicates that after the fast wave steepens the shock propagates with a constant velocity in this region. Therefore, for the case of C2 the fast wave steepens to a shock fastest from the investigated cases and the shock thus forms at the lowest heights in the corona (Figure~\ref{fig:results/sheath}(c)). For C3 in turn, the wave decelerates more slowly and the shock forms at a greater height. 

For run C4 the higher injection velocity does not have a direct influence on the steepening rate (Equation~\ref{eq:steepening-rate}), and the shock starts at approximately the same height as for C1 in Figure~\ref{fig:results/sheath}(c). However, because the FR in C4 propagates much faster through the corona than for C1 (due to the higher injection velocity), it drives the shock faster at the CME nose, and so the difference in the shock location between C4 and C1 increases as the simulation progresses. 
Finally, the cases C5 and C6 show that the frequency of the upstream Alfv\'enic perturbations seem to affect the speed of the CME shock. We note that this dependence of the shock speed on the Alfv\'en wave frequency is independent of the grid resolution (Appendix~\ref{appendix:shock-compression-aw}).  
In Section~\ref{subsec:methods/AW}, we showed that Alfv\'enic waves in the solar wind in our simulation could not steepen to form shocks themselves. Only the interaction of the Alfv\'en waves with the shock, and the initially propagating fast MHD wave prior to shock formation may alter the shock speed. It should be noted that this effect of the Alfv\'en wave frequency affecting the shock formation is shown for a quasi-parallel shock in this simulation. In the case of a perpendicular shock, previous studies~\citep{lu2009interaction} only indicated structural modifications at the shock front without an influence on the propagation speed.

Thus, the propagation velocity of the CME shock depends initially on the effect of the wave steepening followed by the FR driving it further. The FR itself propagates based on the injection velocity and momentum contributing to the resulting force imbalance at onset. This variation of how different parameters affect the shock and FR location separately causes differences in the CME sheath extent.


\section{Conclusion}     \label{sec:conclusion}
This study presents the interaction between small amplitude Alfv\'enic fluctuations and a CME in the low corona using 2.5D time-dependent MHD simulations. The fluctuations in the quasi-steady solar wind are linearly polarised and monochromatic in frequency. They are injected using time-dependent boundary conditions in the low corona.

In Section~\ref{subsec:methods/AW} we described the linear evolution of the injected Alfv\'en waves without decaying into compressive and reflected wave modes. In this scenario we found that the CME sheath would consist of short-wavelength waves that are compressed by the shock and long-wavelength waves transmitted in the initial quasi-perpendicular phase of the CME expansion, which were modified in wavelength by the CME shock passage. The Alfv\'en waves downstream of the CME shock were inhibited close to the FR due to non-radial flows. While this result was obtained for a 2D simulation, we can extrapolate this argument into a higher dimension. In a 3D case we would observe non-radial flows in $\phi$ at the CME flanks as well (in the same manner as we do for $\theta$). Thus, we might expect the CME sheath fluctuations to consist of short-wavelength components based on the non-radial flows present in each direction. Due to the importance of the CME sheath structure in influencing the fluctuations present in this region, we investigated the formation of the sheath in Section~\ref{sec:cme-sheath}. We found the CME-driven shock to be formed due to wave steepening, with an additional constraint on the frequency of the fluctuations present in the system. At the same time the FR evolution is unaffected by the frequency of the fluctuations. In the discussion presented in this study, we do not address the Alfv\'en waves generated by the magnetic recconection~\citep{cranmer2018low, lynch2014interchange} inside the CME sheath. We cannot capture these additional waves in this simulation as they require a much finer simulation grid. The properties of such reconnection-driven Alfv\'en waves depend on the rate of reconnection, plasma $\beta$, and magnetic field strength~\citep{kigure2010generation}. A complete discussion of the impact of these waves on the observed properties of fluctuations inside CME sheaths would require further study. The results presented in this study are thus in the context of the shock transmission of already-present solar wind fluctuations. Therefore, the applicability of these results is valid close to the CME shock when compared with spacecraft observations. 


A primary result of this study is the transmission of the upstream Alfv\'en waves based on the upstream solar wind conditions in the frame of reference of the CME shock (Section~\ref{subsec:shock-transmission}). This transmission process naturally generates sunward-propagating Alfv\'en waves, with the compression of the upstream anti-sunward propagating waves varying in the latitudinal direction ($\theta$) due to the varying shock speeds. This indicates that Alfv\'enic fluctuations have only anti-sunward components upstream, and both sunward and anti-sunward components downstream due to their interaction with the CME shock. This behaviour has been observed across CME shocks~\citep{good2020cross}. Additionally, the properties of the downstream Alfv\'enic fluctuations depend on their relative distance to the CME shock; locations closer to the shock contain more compressed upstream waves. This might indicate varying spectral slopes in the near-shock, mid-sheath, and near-FR regions of the CME sheath~\citep{kilpua2020magnetic}. In Section~\ref{sec:cme-sheath} we observed that the Alfv\'en wave frequency affects the shock velocity. This requires further study, as previous studies investigating this interaction for perpendicular shocks found no appreciable differences in shock speeds. Thus, the result presented in this study might be a feature of the quasi-parallel CME shock.


\begin{acknowledgements} 
The work has been supported by the Finnish Centre of Excellence in Research on Sustainable Space (FORESAIL). This is a project under the Academy of Finland, and this research has been supported by the European Research Council (SolMAG; grant no. 724391) as well as Academy of Finland project SWATCH (343581). The authors are also grateful to the anonymous referee for their constructive input during the peer review process.
\end{acknowledgements}

\bibliographystyle{aa-note} 
\bibliography{example}      

\clearpage
\begin{appendix}
    
\section{Shock compressed Alfv\'en waves}
\label{appendix:shock-compression-aw}
In this Appendix we present a modified simulation of the nominal run (case C1 in Table~\ref{tab:cme-list}) discussed in Section~\ref{sec:results} where we improve the resolution between $1.1~R_\odot$ and $10~R_\odot$ to capture the compression of upstream Alfv\'en waves due to the shock that are not resolved in Section~\ref{subsec:shock-transmission}. Here we detail the methodology for introducing the modified resolution, and report the expected propagation of Alfv\'en waves based on the discussion presented in Section~\ref{sec:results}. Therefore, the results presented in this Appendix validate the approach presented in this manuscript where we utilise a reduced simulation resolution, while still obtaining the proper physics.

\subsection{Modifying the grid resolution}
As described in Section~\ref{sec:methodology}, the simulation is performed in three steps:
\begin{enumerate}
    \item generating a steady-state solar wind solution without fluctuations;
    \item generating a quasi-steady state wind by injecting Alfv\'en waves;
    \item introducing the CME.
\end{enumerate}

\begin{figure}[ht]
        \centering
        \includegraphics[width=0.5\textwidth]{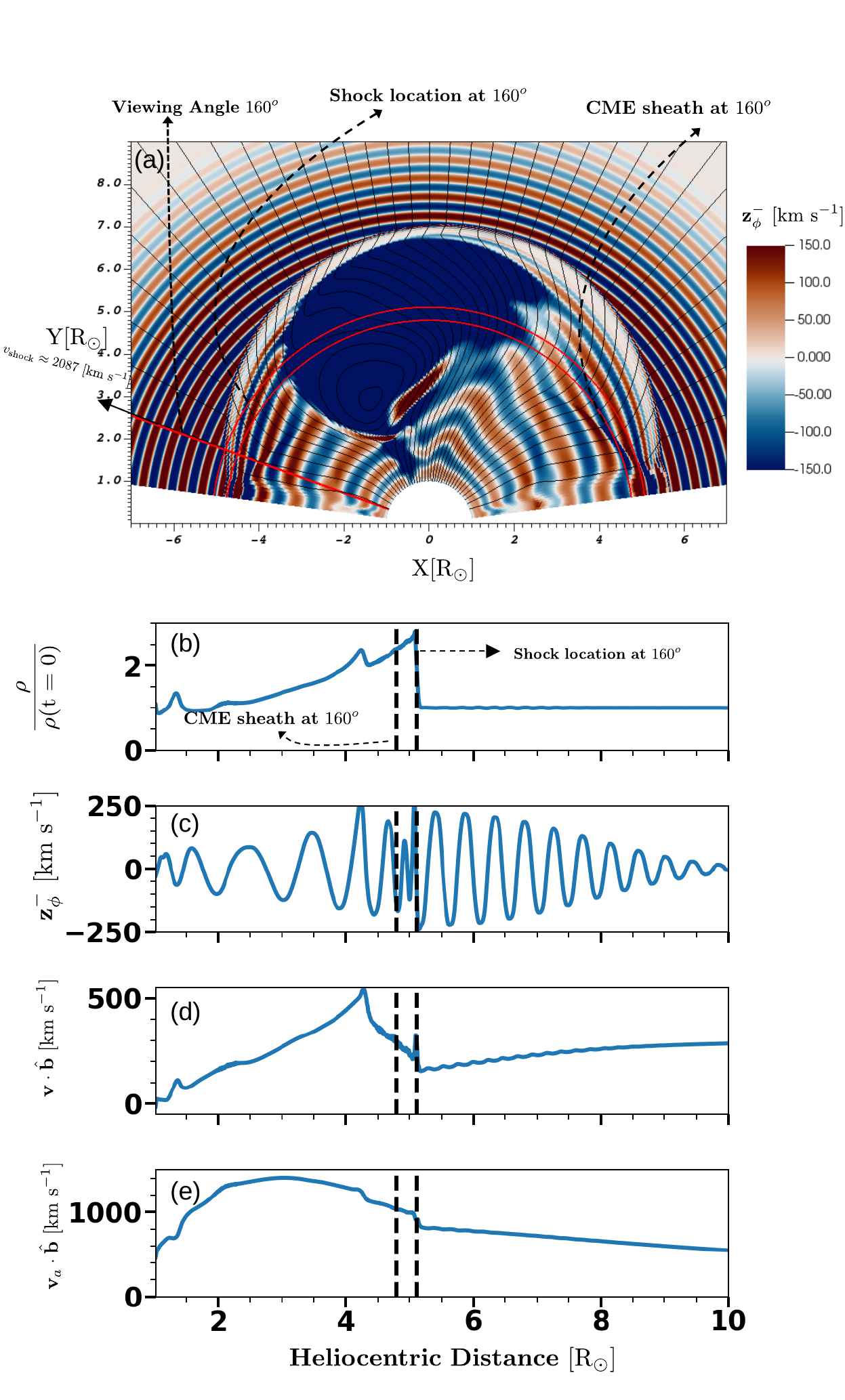}
        \caption{Shock at the CME flank for the modified resolution simulation run. This figure is similar to Figure~\ref{fig:compression}, but  with a modified grid (higher resolution) between $1.1~R_\odot$ and $10~R_\odot$. Panel (a) shows a simulation snapshot at $t=25~$min of the anti-sunward Els\"asser variable $\mathbf{z_\phi^-}$ with annotations describing the viewing angle along $160^\circ$, the shock location, the approximate beginning of the CME sheath, and the approximate shock velocity $v_\mathrm{shock}$. Panels (b) and (c) show the density compression and $\mathbf{z_\phi^-}$ along the viewing angle, respectively. Panels (d) and (e) present the fluid velocity and Alfv\'en speed along the direction of the background magnetic field.}
        \label{fig:compression-modGrid}
\end{figure}

A reduced resolution with $500$ grid cells logarithmically spaced in the radial direction results in numerical damping of the Alfv\'en waves in the corona, and inhibits the generation of density fluctuations and alternate Alfv\'en wave modes (Section~\ref{subsec:methods/AW}). Thus, after obtaining a quasi-steady wind with a reduced resolution, we modify the simulation grid to have $4000$ equally spaced cells between $1.1~R_\odot$ and $10~R_\odot$ solar radii, while preserving the original grid below $1.1~R_\odot$ and beyond $10~R_\odot$. This higher resolution in the region of interest where the Alfv\'en waves are present in the simulation enables us to capture the compression at the CME shock. The choice of $4000$ cells was made after determining that the solar wind behaviour is unchanged for $3000$ and $3500$ grid cells as well. 

Figure~\ref{fig:compression-modGrid} illustrates the effect of the modified grid on the Alfv\'en wave transmission by presenting a figure similar to Figure~\ref{fig:compression}. In panels (a) and (c) we observe that Alfv\'en waves, seen as crests and troughs of the $z^-_\phi$ Els\"asser variable, are of a higher amplitude than in Figure~\ref{fig:compression}, due to a lower level of numerical damping. However, the CME shock and the start of the longer wavelength CME sheath region is seen to occur at the same location as in the lower-resolution simulation, as annotated in the figure. The primary difference between Figure~\ref{fig:compression-modGrid} and Figure~\ref{fig:compression} is the shock compressed region in panel (c) between the vertical annotations. In this region we observe the upstream waves compressed in frequency (as described in Section~\ref{subsec:shock-transmission}), with the modified grid capturing one wavelength of this compressed Alfv\'en wave. However, after this compressed region the Alfv\'en waves observed further downstream are also of longer wavelengths, as in the lower-resolution runs, as the Alfv\'en speed increases again. To verify whether the grid resolution captures the entire compressed wave, we plot the flow speed ($\mathbf{v\cdot}\hat{\mathbf{b}}$), and the Alfv\'en speed along the background field ($\mathbf{v}_a\cdot\hat{\mathbf{b}}$) in Figure~\ref{fig:compression-modGrid}(d),~(e). Assuming the shock normal to be exactly in the direction of the upstream magnetic field (in the radial direction), and by estimating the shock speed between $t = 6.25~$min and $t = 31.25~$min as $\approx 2087~\mathrm{km~s^{-1}}$, we obtain the Alfv\'en Mach number in the shock frame to be $\approx 2.45$ and $\approx 1.78$ in the upstream and downstream, respectively. Additionally, taking the gas compression ratio to be $\approx 2$ (panel (d)), we can estimate the downstream wavelength (Equation~\ref{eq:anti-sunward-kvec}) to be three times smaller than the upstream wavelength, which can be resolved by the higher-resolution grid. It should be noted that the average shock velocity with the modified grid ($v_\mathrm{shock} \approx 2087~\mathrm{km~s^{-1}}$) is marginally higher than for the original grid ($v_\mathrm{shock} \approx 2078~\mathrm{km~s^{-1}}$), possibly due to the modified grid better resolving the interactions of the wave with the shock.

Finally, the effect of the Alfv\'en waves on the propagation of the CME is discussed in Section~\ref{sec:cme-sheath}. The FR propagates based on its force imbalance, and does not present a significant dependence on the Alfv\'en wave frequency. In contrast, the shock velocity increases with the wave frequency. To verify whether such a result is dependant on the grid resolution, we performed a similar analysis as in Section~\ref{sec:cme-sheath} to check the nominal case C1 against the simulation with the modified resolution as in the Appendix. The results showed that the Alfv\'en wave has no further effect on the CME shock velocity with increased resolution.

\subsection{Alfv\'en wave transmission across the shock}
\begin{figure}[ht]
        \centering
        \includegraphics[width=0.5\textwidth]{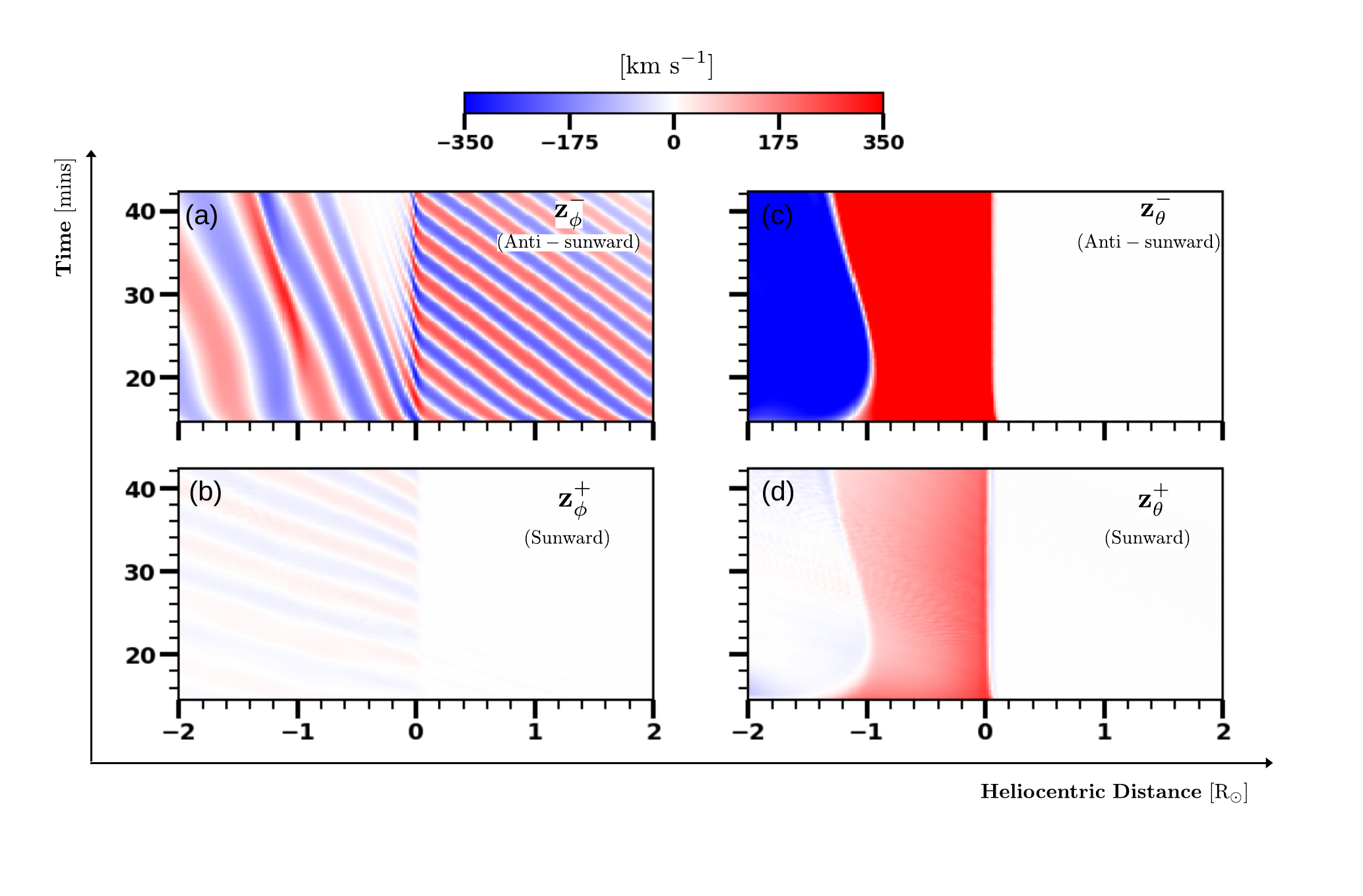}
        \caption{Evolution of the Els\"asser variables at the CME flank for the modified simulation grid. Similar to Figure~\ref{fig:results/elss-3mHz}, but using a modified grid (higher resolution) between $1.1~R_\odot$ and $10~R_\odot$. The x-axis similarly denotes the shock neighbourhood in units of $R_\odot$; the shock is centred at $x=0$.}
        \label{fig:appendix/aroundMod-160}
\end{figure}

\begin{figure}[ht]
        \centering
        \includegraphics[width=0.5\textwidth]{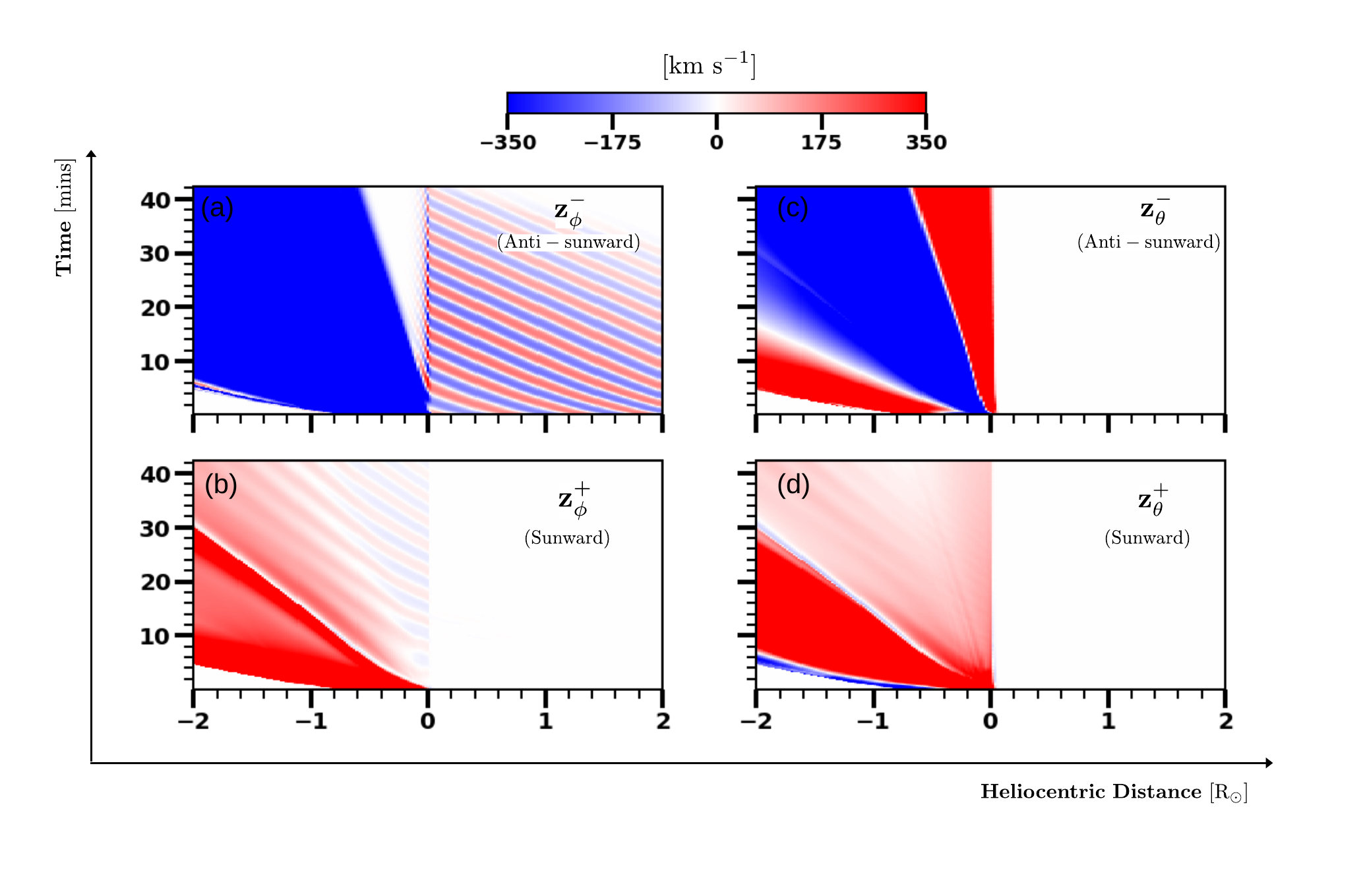}
        \caption{Evolution of the Els\"asser variables at the CME nose for the modified simulation grid, similar to Figure~\ref{fig:appendix/aroundMod-160}, but with a viewing angle of $105^\circ$. The x-axis similarly denotes the shock neighbourhood in units of $R_\odot$; the shock is centred at $x=0$.}
        \label{fig:appendix/aroundMod-105}
\end{figure}

The transmission of the upstream Alfv\'en wave occurs based on the gas compression ratio across the shock and the upstream Alfv\'en Mach number in the shock frame. To illustrate the compressed wave further, Figures~\ref{fig:appendix/aroundMod-160} and \ref{fig:appendix/aroundMod-105} are similar to Figure~\ref{fig:results/elss-3mHz}, but use the modified high-resolution grid, and are presented for viewing angles of $160^\circ$ and $105^\circ$.

In Figure~\ref{fig:appendix/aroundMod-160} (comparable to Figure~\ref{fig:results/elss-3mHz}) we see the upstream waves in panel (a) incident to the shock with the far downstream region ($x < 0$) showing the long-wavelength Alfv\'en waves. The downstream region also shows an enhancement of the Alfv\'en wave amplitudes beyond $\mathrm{t} = 20~\mathrm{min}$. We now also observe the compressed waves near the shock, which are amplified, but quickly dissipate as the shock propagates further. This dissipation and subsequent enhancement of the downstream waves seems to indicate steepening of the compressed waves. The Els\"asser variables in $\theta$ remain as presented in Section~\ref{subsec:fluctuations-around-shock}.

For the viewing angle of $105^\circ$ (CME head-on) in Figure~\ref{fig:appendix/aroundMod-105} we see a similar compression of the upstream waves and subsequent dissipation as the shock progresses further. The far downstream region shows the expected large flow due to the presence of the FR magnetic field. Panels (b), (c), and (d) show behaviour similar to that in Figure~\ref{fig:results/elss-3mHz-105}. 

Therefore, the upstream Alfv\'en waves are compressed by the CME shock and further interact non-linearly with each other to dissipate and steepen the downstream waves. The high-resolution grid captures the expected behaviour of the waves, as described in Section~\ref{subsec:shock-transmission}, with an additional effect based on the wave dissipation as the CME propagates beyond $10~\mathrm{R_\odot}$. The resolution does not appear to have any effect on the CME dynamics.
\end{appendix}
\end{document}